\documentclass[sigconf, nonacm]{acmart}

\usepackage{amsmath}
\usepackage[ruled,vlined]{algorithm2e}
\usepackage{multirow}
\usepackage{subfigure}
\usepackage{graphicx}
\usepackage{bm}

\newcommand{\xkliao}{\textcolor{blue}}
\newcommand{\gc}{\textcolor{black}}
\newtheorem{property}{Property}[section]
\newtheorem{problem}{Problem}
\newtheorem{theorem}{Theorem}[section]
\newtheorem{definition}{Definition}[section]
\newtheorem{example}{Example}[section]
\newcommand{\gcc}{\textcolor{blue}}

\newcommand\vldbdoi{XX.XX/XXX.XX}
\newcommand\vldbpages{XXX-XXX}
\newcommand\vldbvolume{14}
\newcommand\vldbissue{1}
\newcommand\vldbyear{2020}
\newcommand\vldbauthors{\authors}
\newcommand\vldbtitle{\shorttitle}
\newcommand\vldbavailabilityurl{URL_TO_YOUR_ARTIFACTS}
\newcommand\vldbpagestyle{plain}

\newcommand{\stitle}[1]{\vspace{1ex} \noindent{\bf #1}}

\newcommand{\kw}[1]{{\ensuremath {\mathsf{#1}}}\xspace}

\newcommand{\SC}{\kw{SC}\xspace}

\begin{document}
\title{Distributed D-core Decomposition over Large Directed Graphs}

\author{Xuankun Liao}
\authornotemark[1]
\affiliation{%
  \institution{Hong Kong Baptist University}
  \streetaddress{ }
  \city{ }
  \state{ }
  \postcode{ }
}
\email{xkliao@comp.hkbu.edu.hk}

\author{Qing Liu}
\authornote{These authors have contributed equally to this work.}
\affiliation{%
  \institution{Hong Kong Baptist University}
  \streetaddress{ }
  \city{ }
  \state{ }
  \postcode{ }
}
\email{qingliu@comp.hkbu.edu.hk}

\author{Jiaxin Jiang}
\affiliation{%
  \institution{Hong Kong Baptist University}
  \streetaddress{ }
  \city{ }
  \state{ }
  \postcode{ }
}
\email{jxjian@comp.hkbu.edu.hk}

\author{Xin Huang}
\affiliation{%
  \institution{Hong Kong Baptist University}
  \streetaddress{ }
  \city{ }
  \state{ }
  \postcode{ }
}
\email{xinhuang@comp.hkbu.edu.hk}

\author{Jianliang Xu}
\affiliation{%
  \institution{Hong Kong Baptist University}
  \streetaddress{ }
  \city{ }
  \state{ }
  \postcode{ }
}
\email{xujl@comp.hkbu.edu.hk}

\author{Byron Choi}
\affiliation{%
  \institution{Hong Kong Baptist University}
  \streetaddress{ }
  \city{ }
  \state{ }
  \postcode{ }
}
\email{bchoi@comp.hkbu.edu.hk}


%
%
%
%

\begin{abstract}
Given a directed graph $G$ and integers $k$ and $l$, a D-core is the maximal subgraph $H \subseteq G$ such that for every vertex of $H$, its in-degree and out-degree are no smaller than $k$ and $l$, respectively. For a directed graph  $G$, the problem of D-core decomposition aims to compute the non-empty D-cores for all possible values of $k$ and $l$.
In the literature, several \emph{peeling-based} algorithms have been proposed to handle D-core decomposition.
However, the peeling-based algorithms that work in a sequential fashion and require global graph information during processing are mainly designed for \emph{centralized} settings, which cannot handle large-scale graphs efficiently in distributed settings.
Motivated by this, we study the \emph{distributed} D-core decomposition problem in this paper. We start by defining a concept called \emph{anchored coreness}, based on which we propose a new H-index-based algorithm for distributed D-core decomposition. Furthermore, we devise a novel concept, namely  \emph{skyline coreness}, and show that the D-core decomposition problem is equivalent to the computation of skyline corenesses for all vertices. We design an efficient D-index to compute the skyline corenesses distributedly.
We implement the proposed algorithms under both vertex-centric and block-centric distributed graph processing frameworks.
Moreover, we theoretically analyze the algorithm and message complexities.  Extensive experiments on large real-world graphs with billions of edges demonstrate the efficiency of the proposed algorithms in terms of both the running time and communication overhead. 
\end{abstract}

\maketitle

\pagestyle{\vldbpagestyle}
\begingroup\small\noindent\raggedright\textbf{PVLDB Reference Format:}\\
\vldbauthors. \vldbtitle. PVLDB, \vldbvolume(\vldbissue): \vldbpages, \vldbyear.\\
\href{https://doi.org/\vldbdoi}{doi:\vldbdoi}
\endgroup
\begingroup
\renewcommand\thefootnote{}\footnote{\noindent
This work is licensed under the Creative Commons BY-NC-ND 4.0 International License. Visit \url{https://creativecommons.org/licenses/by-nc-nd/4.0/} to view a copy of this license. For any use beyond those covered by this license, obtain permission by emailing \href{mailto:info@vldb.org}{info@vldb.org}. Copyright is held by the owner/author(s). Publication rights licensed to the VLDB Endowment. \\
\raggedright Proceedings of the VLDB Endowment, Vol. \vldbvolume, No. \vldbissue\ %
ISSN 2150-8097. \\
\href{https://doi.org/\vldbdoi}{doi:\vldbdoi} \\
}\addtocounter{footnote}{-1}\endgroup


\section{Introduction}
\label{sec:intro}

\begin{figure}[t]
\centering
\subfigure[A directed graph $G$]{
\label{fig:1a}
\includegraphics[width = 0.2\textwidth]{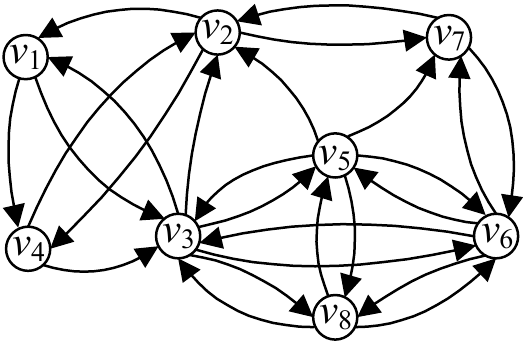}}
\subfigure[All non-empty D-cores]{
\label{fig:1b}
\includegraphics[width = 0.25\textwidth]{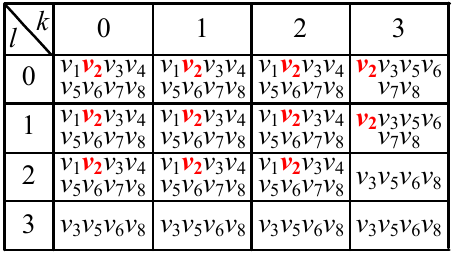}}
\vspace{-0.5cm}
\caption{An example of D-core decomposition on $G$}
\vspace{-0.15in}
\label{fig:introexample}
\end{figure}

Graph is a widely used data structure to depict entities and their relationships. In a directed graph, edges have directions to represent links from one vertex to another, 
which has 
many real-life applications,
e.g., the \emph{following} relationship in online social networks such as Twitter, the \emph{money flow} in financial networks, the \emph{traffic route} in road networks, and the \emph{message forwarding path} in communication networks. Among many graph analysis algorithms, cohesive subgraph analysis is to discover densely connected subgraphs under a cohesive subgraph model. A well-known model used for undirected graphs is $k$-core, which requires every vertex in the subgraph to have at least $k$ neighbors~\cite{k-core-definition}. As a directed version of $k$-core, $D$-core, a.k.a. $(k, l)$-core, is the maximal directed subgraph such that every vertex has at least $k$ in-neighbors and $l$ out-neighbors within this subgraph~\cite{dcore}. For example, in Figure~\ref{fig:1a}, the whole directed graph $G$ is a (2, 2)-core since every vertex has an in-degree of at least 2 and an out-degree of at least 2.  



As a foundation of D-core discovery, the problem of D-core decomposition aims to compute the non-empty D-cores of a directed graph for all possible values of $k$ and $l$.
D-core decomposition has a number of applications. It has been used to \emph{build coreness-based indexes for speeding up community search}~\cite{dcore-cs,chen2020efficient}, to \emph{measure influence in social networks}~\cite{garcia2017understanding}, to \emph{evaluate graph collaboration features of communities}~\cite{dcore}, to \emph{visualize and characterize complex networks}~\cite{distributedk-core}, and to \emph{discover hubs and authorities} of directed networks~\cite{soldano2017hub}. \gcc{For example, based on the D-core decomposition results, we can index a graph's vertices by their corenesses using a table~\cite{dcore-cs} or D-Forest~\cite{chen2020efficient}; then, D-core-based community search can be accelerated by looking up the table or D-Forest directly, instead of performing the search from scratch.}
In the literature, \emph{peeling-based} algorithms have been proposed for D-core decomposition \gcc{in centralized settings}~\cite{dcore,dcore-cs}.  They work in a sequential fashion to remove disqualified vertices one by one from a graph.
That is, they first determine all possible values of $k$ (i.e., from 0 to the maximum in-degree of the graph). Next, for each value $k$, they compute $(k, l)$-cores for all possible values of $l$ by iteratively deleting the vertices with the smallest out-degree.
Figure~\ref{fig:1b} shows the  D-core decomposition results of $G$ for different $k$ and $l$ values.

In this paper, we study the problem of D-core decomposition in distributed settings, where the input graph $G$ is stored on a collection of machines and each machine holds only a partial subgraph of $G$. \gcc{The motivation is two-folded. First, due to the large size of graph data, D-core decomposition necessitates huge memory space, which may exceed the capacity of a single machine. For example, the existing algorithms could not work for billion-scale graphs due to excessive memory space costs~\cite{dcore-cs}. Second, in practical applications, many large graphs are inherently distributed over a collection of machines, making distributed processing a natural solution~\cite{siamcompLinial92, kcoredecompositiondistributeddynamic,distributedk-core,bigdataconfMandalH17}.}

However, the existing peeling-based algorithms are not efficient when extended to distributed settings. In particular, when computing the $(k, l)$-cores for a given $k$, \gcc{the algorithms need to iteratively find the vertices with the smallest out-degree to delete and then update the out-degrees for the remaining vertices, until the graph becomes empty. This process (i) is not parallelizable since the update of out-degrees in each iteration depends on the vertices deletion in the previous iteration and (ii) entails expensive network communications since it needs global graph information.}

To address these issues, we design new distributed D-core decomposition algorithms by exploiting the relationships between a vertex and its neighbors.
First, inspired by the notion of \emph{k-list}~\cite{dcore-cs}, we propose an \emph{anchored coreness}-based algorithm. Specifically, for a vertex $v$, if we fix the value of $k_v$, we can compute the maximum value of $l_v$ such that  $v$ is contained in the $(k_v, l_v)$-core. We call this pair $(k_v, l_v)$ an anchored coreness of $v$. For example, for vertex $v_2$ in Figure~\ref{fig:introexample}, when $k_{v_2}=0$, the maximum value of $l_{v_2}$ is 2 since $v_2 \in (0, 2)$-core but $v_2 \notin (0, 3)$-core. Hence, $(0, 2)$ is an anchored coreness of $v_2$. The other anchored corenesses of $v_2$ are $(1, 2)$, $(2, 2)$, and $(3, 1)$. Once we have computed the anchored corenesses for every vertex, we can easily derive the D-cores from these anchored corenesses. Specifically, given integers $k$ and $l$, the $(k, l)$-core consists of the vertices whose anchored coreness $(k_v, l_v)$ satisfies $k_v = k$ and $l_v \geq l$. Thus, the problem of distributed D-core decomposition is equivalent to computing the anchored corenesses in a distributed way. To do so,
we first exploit the in-degree relationship between a vertex and its in-neighbors and define an \emph{in-H-index}, based on which we compute the maximum value of $k$ for each vertex. Then, we study the property of  $(k,0)$-core and define an \emph{out-H-index}. On the basis of that, for each possible value of $k$ with respect to a vertex, we iteratively compute the corresponding upper bound of $l$ simultaneously. Finally, we utilize the definition of D-core to iteratively refine all the upper bounds to obtain the anchored corenesses of all vertices.

\gc{Note that the anchored coreness-based algorithm first fixes one dimension and then computes the anchored corenesses for the other dimension, which may lead to suboptimal performance.}
To improve performance, 
we further propose a novel concept, called \emph{skyline coreness}, and develop a \emph{skyline coreness}-based algorithm. Specifically, we say the pair $(k_v, l_v)$ is a skyline coreness of a vertex $v$, if there is no other pair $(k_v', l_v')$ such that $k_v' \geq k_v$, $l_v' \geq l_v$, and $v \in (k_v', l_v')$-core.
For example, in Figure~\ref{fig:introexample}, the skyline corenesses of $v_2$ are $\{(2, 2), (3, 1)\}$. Compared with anchored corenesses, a vertex's skyline corenesses contain fewer pairs of $(k_v, l_v)$. Nevertheless, based on the skyline corenesses, we can still easily find all the D-cores containing the corresponding vertex. To be specific, if $(k_v, l_v)$ is a skyline coreness of $v$, then $v$ is also in the $(k, l)$-cores with $k \leq k_v$ and $l \leq l_v$. 
The basic idea of the skyline coreness-based algorithm is to use neighbors' skyline corenesses to iteratively estimate the skyline corenesses of each vertex. To this end,
we define a new index, called \emph{D-index}, for each vertex based on the following unique property of skyline corenesses. If $(k_v, l_v)$ is one of the skyline corenesses of a vertex $v$, we have (i) $v$ has at least $k_v$ in-neighbors such that each of these in-neighbors, $v_i$, has a skyline coreness $(k_{v_i}', l_{v_i}')$ satisfying $k_{v_i}' \geq k_v$ and $l_{v_i}' \geq l_v$; and (ii) $v$ has at least $l_v$ out-neighbors such that each of these out-neighbors, $v_j$,  has a skyline coreness $(k_{v_j}'', l_{v_j}'')$ satisfying $k_{v_j}'' \geq k_v$ and $l_{v_j}'' \geq l_v$. 
With this property, we design a distributed algorithm to iteratively compute the D-index for each vertex with its neighbors' D-indexes. \gc{To deal with the combinatorial blow-ups in the computation of D-indexes, we further develop three optimization strategies to improve efficiency.}

We implement our algorithms under two well-known distributed graph processing frameworks, i.e., vertex-centric~\cite{pregel,giraph,gps,graphlab}  and block-centric~\cite{giraph++,blogel,fan2017grape}. Empirical results on small graphs demonstrate that
   our algorithms run faster than the peeling-based algorithm by up to 3 orders of magnitude.
   For larger graphs with more than 50 million edges, the peeling-based algorithm cannot finish within 5 days, while our algorithms can finish within 1 hour for most datasets.  \gcc{Moreover, our proposed algorithms require less than 100 rounds to converge for most datasets, and more than $90\%$ vertices can converge within 10 rounds.} 


This paper's main contributions are summarized as follows:
\begin{itemize}
    \item For the first time in the literature, we study the problem of distributed D-core decomposition over large directed graphs.
    \item We develop an anchored coreness-based distributed algorithm using well-defined in-H-index and out-H-index. To efficiently compute the anchored corenesses, we propose tight upper bounds that \gc{can be iteratively refined to exact anchored corenesses with reduced network communications}.

    \item We further propose a novel concept of skyline coreness and show that the problem is equivalent to the computation of skyline corenesses for all vertices. A new two-dimensional D-index that unifies the in- and out-neighbor relationships, together with three optimization strategies, is designed to  compute the skyline corenesses distributedly. %
    \item 
        Both theoretical analysis and empirical evaluation validate the efficiency of our algorithms  for distributed D-core decomposition.
\end{itemize}

The rest of the paper is organized as follows. Section~\ref{Sec:Relatedwork} reviews related work. Section~\ref{Sec:ProblemFormulation} formally defines the problem. Sections~\ref{Sec:ACC} and~\ref{Sec:SCC} propose two distributed algorithms for computing anchored coreness and skyline coreness, respectively. Experimental results are reported in Section~\ref{Sec:experiment}. Finally, Section~\ref{Sec:conclusion} concludes the paper.

\section{Related Work}
\label{Sec:Relatedwork}

In this section, we review the related work in two aspects, i.e., \emph{core decomposition} and \emph{distributed graph computation}.

\stitle{Core Decomposition.} As a well-known dense subgraph model, a $k$-core is the maximal  subgraph of an undirected graph such that every vertex has at least $k$ neighbors within this subgraph~\cite{k-core-definition}. The core decomposition task aims at finding the $k$-cores for all possible values of $k$ in a graph. Many efficient algorithms have been proposed to handle core decomposition over an undirected graph, such as peeling-based algorithms~\cite{kcore,k-core-efficient,kcoredecompositiononsinglePC}, disk-based algorithm~\cite{k-core-efficient,kcoredecompositiononsinglePC},  semi-external algorithm~\cite{icdeWenQZLY16}, streaming algorithms~\cite{kcoredecompositionstreaming,kcoredecompositionincremental},  parallel algorithms~\cite{kcoredecompositionmulticore,kcoredecomposisitonparallelstreaming}, and distributed algorithms~\cite{kcoredecompositiondistributeddynamic,distributedk-core,bigdataconfMandalH17}. \gcc{
It is worth mentioning that the distributed algorithms for $k$-core decomposition~\cite{kcoredecompositiondistributeddynamic,distributedk-core,bigdataconfMandalH17} cannot be used for distributed D-core decomposition. Specifically, the distributed  $k$-core decomposition algorithms use the neighbors' corenesses to estimate a vertex's coreness, where all neighbors are of the same type. For D-core, a vertex's neighbors include in-neighbors and out-neighbors, which affect each other and should be considered simultaneously. If we  consider only one type of neighbors, we cannot get the correct answer.
Inspired by the H-index-based computation for core decomposition~\cite{lu2016h} and nucleus decomposition~\cite{hierarchical-dense-subgrah}, we apply a similar idea in the design of distributed algorithms.  Nevertheless, our technical novelty lies in the non-trivial extension of H-index from \emph{one-dimensional undirected coreness} to \emph{two-dimensional anchored/skyline coreness}, which needs to consider the computations of in-degrees and out-degrees simultaneously in a unified way. }

In addition, core decomposition has  been studied for different types of networks, such as weighted graphs~\cite{kcoredecompositionweightedgraph,coredecompositionandmaintenceweighted}, uncertain graphs~\cite{coredecompositionuncertain,icdePengZZLQ18}, bipartite graphs~\cite{wwwLiuYLQZZ19}, temporal graphs~\cite{tkddGalimbertiCBBCG21,bigdataconfWuCLKHYW15}, 
and heterogeneous information networks~\cite{pvldbFangYZLC20}. Recently, a new problem of distance-generalized core decomposition has been studied by considering vertices' $h$-hop connectivity~\cite{khcoredecompositon,pvldbqingliu}. 
\gcc{
Note that a directed graph can be viewed as a bipartite graph.   After transforming  a directed graph to a bipartite graph, the $(k,l)$-core in the directed graph has a corresponding  $(\alpha,\beta)$-core in the bipartite graph~\cite{wwwLiuYLQZZ19}, but not vice versa. Therefore, the problems of $(k,l)$-core decomposition and $(\alpha,\beta)$-core decomposition are not equivalent, 
 and $(\alpha,\beta)$-core decomposition algorithms cannot be used in our work.}
\stitle{Distributed Graph Computation.} In the literature, there exist various distributed graph computing models and systems to support big graph analytics.
Among them, the \emph{vertex-centric} framework~\cite{pregel,graphlab,csurMcCuneWM15} and the \emph{block-centric} framework~\cite{giraph++,blogel} are  two most popular frameworks. 

The vertex-centric framework assumes that each vertex is associated with one computing node and communication occurs through edges. The workflow of the vertex-centric framework consists of a set of synchronous supersteps.  Within each superstep, the vertices  execute a user-defined function asynchronously after receiving messages from their neighbors. If a vertex does not receive any message, it will be marked as inactive.  The framework stops once all vertices become inactive. Typical vertex-centric systems include Pregel~\cite{pregel}, Giraph~\cite{giraph}, GPS~\cite{gps}, and GraphLab~\cite{graphlab}.
For the block-centric framework, one computing node stores the vertices within a block together and communication occurs between blocks after the computation within a block reaches convergence.  Compared with the  vertex-centric framework, the block-centric framework can reduce the network traffic and better balance the workload among nodes. Distributed graph processing systems such as Giraph++~\cite{giraph++}, Blogel~\cite{blogel}, and GRAPE~\cite{fan2017grape} belong to the block-centric framework.

Note that in this paper we mainly focus on algorithm designs for distributed D-core decomposition. To demonstrate the flexibility of our proposed algorithms, we implement them for performance evaluation in both vertex-centric and block-centric frameworks.

\section{Problem Formulation}
\label{Sec:ProblemFormulation}

In this paper, we consider a directed, unweighted simple graph $G=(V_G, E_G)$, where $V_G$ and $E_G$ are the set of vertices and edges, respectively. Each edge $e\in E_G$ has a direction.
For an edge $e=\langle u, v\rangle \in E_G$, we say $u$ is an in-neighbor of $v$ and $v$ is an out-neighbor of $u$. Correspondingly, $N^{in}_G(v)$ and $N^{out}_G(v)$ are respectively denoted as the in-neighbor set and out-neighbor set of a vertex $v$ in $G$. We define three kinds of degrees for a vertex $v$ as follows: (1) the in-degree is the number of $v$'s in-neighbors in $G$, i.e., $deg^{in}_G(v) = |N^{in}_G(v)|$; (2) the out-degree is the number of $v$'s out-neighbors in $G$, i.e., $deg^{out}_G(v) = |N^{out}_G(v)|$; (3) the degree  is the sum of its in-degree and out-degree, i.e., $deg_G(v) = deg^{in}_G(v)+ deg^{out}_G(v)$.
Based on the 
in-degree and out-degree, we give a definition of D-core as follows.






\begin{definition}
\textbf {D-core \cite{dcore}.} Given a directed graph $G=(V_G, E_G)$ and two integers $k$ and $l$, a D-core of $G$, also denoted as $(k, l)$-core, is the maximal subgraph $H =(V_H, E_H) \subseteq G$ such that $\forall v \in V_H$, $deg_H^{in}(v) \geq k$ and $deg_H^{out}(v) \geq l$.
\label{def:dcore}
\end{definition}

According to Definition~\ref{def:dcore},  a D-core should satisfy both the degree constraints and the size constraint. The degree constraints ensure the cohesiveness of D-core in terms of in-degree and out-degree. The size constraint guarantees the uniqueness of the D-core, i.e., for a specific pair of $(k, l)$, there exists at most one D-core in $G$. Moreover, D-core has a \emph{partial nesting} property as follows.

\begin{property}
\textbf {Partial Nesting.}  Given two D-cores, $(k_{1}, l_{1})$-core $D_1$ and $(k_{2}, l_{2})$-core $D_2$, $D_1$ is nested in $D_2$ (i.e., $D_1 \subseteq D_2$) if $k_{1} \geq k_{2}$ and $l_{1} \geq l_{2}$.
Note that  if  $k_{1} \geq k_{2}$ and $l_{1} < l_{2}$, or $k_{1} < k_{2}$ and $l_{1} \geq l_{2}$, $D_1$ and $D_2$ may be not nested in each other.  
\label{pro.nest}
\end{property}

\vspace{-0.1in}

\begin{figure}[tp]
\vspace{-0.15in}
\centering
\includegraphics[width = 0.25\textwidth]{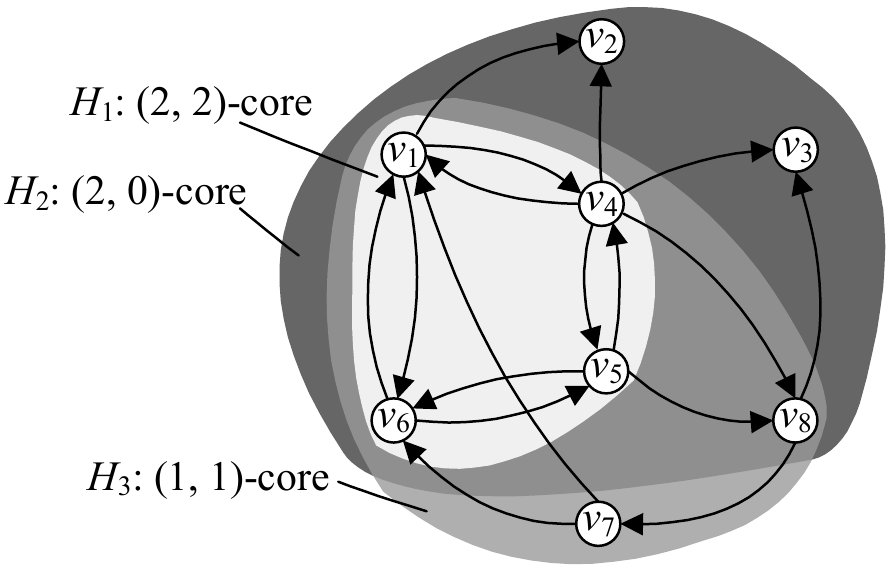}
\vspace{-0.1in}
\caption{D-core}
\vspace{-0.1in}
\label{fig:exampleDcore}
\end{figure}

\begin{example}
In Figure~\ref{fig:exampleDcore}, the directed subgraph $H_1$ induced by the vertices $v_1$, $v_4$, $v_5$, and $v_6$ is a $(2, 2)$-core since $\forall v\in V_{H_1}$, $deg^{in}_{H_1}(v) = deg^{out}_{H_1}(v) =2$. Moreover, $H_1$ $\subseteq$ $H_2=(2, 0)$-core, $H_1$ $\subseteq$ $H_3=(1, 1)$-core. On the other hand, $H_2$ $\nsubseteq$ $H_3$ and $H_3$ $\nsubseteq$ $H_2$, due to the non-overlapping vertices $v_2$, $v_3$, and $v_7$.
\end{example}

In this paper, we study the problem of D-core decomposition to find all D-cores of a directed graph $G$ in distributed settings. In-memory algorithms of D-core decomposition have been studied in \cite{dcore,dcore-cs}, assuming that the entire graph and associated structures can fit into the memory of a single machine. To our best knowledge, the problem of distributed D-core decomposition, considering a large graph distributed over a collection of machines, has not been investigated in the literature. We formulate a new problem of distributed D-core decomposition as follows.


\begin{problem}
{\bf (Distributed D-core Decomposition).} Given a directed graph $G=(V_G, E_G)$ that is distributed in a collection of machines $\{M_i:$  a machine  $M_i$ holds a partial subgraph $G_i\subseteq G$, $1\leq i\leq n\}$ where $n\geq2$ and $\cup_{i=1}^{n} G_i = G$, the problem of distributed D-core decomposition is to find all D-cores of $G$ using $n$ machines, i.e., identifying the $(k, l)$-cores with all possible $(k, l)$ pairs.
\label{problem:prob1}
\end{problem}

Consider applying D-core decomposition on $G$ in Figure~\ref{fig:exampleDcore}. We can obtain a total of 9 different D-cores: $(0, 0)$-core $=$ $(1, 0)$-core $=$ $G$; $(0, 1)$-core $=$ $(1, 1)$-core $=$ $H_3$; $(0, 2)$-core $=$ the subgraph of $G$ induced by the vertices in $V_{H_1} \cup \{v_7\}$; $(1, 2)$-core $=$ $(2, 1)$-core $=$ $(2, 2)$-core $=$ $H_1$; $(2, 0)$-core $=$ $H_2$.


In the following two sections, we propose two new distributed algorithms for D-core decomposition. Without loss of generality, we mainly present the algorithms under the vertex-centric framework. At the end of Sections~\ref{Sec:ACC} and \ref{Sec:SCC}, we discuss how to extend our proposed algorithms to the block-centric framework.


\section{Distributed Anchored Coreness-Based Algorithm}
\label{Sec:ACC}
In this section, we first give a definition of anchored coreness, which is useful for D-core decomposition. Then, we present a vertex-centric distributed algorithm for anchored coreness computation. Finally, we analyze the correctness and complexity of our proposed algorithm, and discuss its block-centric extension.

\subsection{Anchored Coreness}

Recall that, in the undirected $k$-core model~\cite{k-core-definition}, every vertex $v$ has a unique value called \emph{coreness}, i.e., the maximum value $k\in \mathbb{N}_0$ such that  $v$ is contained in a non-empty $k$-core.
Similarly, we give a definition of \emph{anchored coreness} for directed graphs as follows.

\begin{definition}\gcc{
\textbf {(Anchored Coreness).} Given a directed graph $G$ and an integer $k$, the anchored coreness of a vertex $v\in V_G$ is a pair $(k, l_{max}(v, k))$, where  $l_{max}(v, k) = \max \limits_{l\in \mathbb{N}_0} \{l  \ | \ \exists (k, l)\text{-core } H\subseteq G \wedge v \in V_H\}$.
The \emph{entire anchored corenesses} of the vertex $v$ are defined as $\Phi(v)=\{(k', l_{max}(v, k'))  \ |  \ 0 \leq k' \leq k_{max}(v)\}$, where $k_{max}(v) = \max \limits_{k''\in \mathbb{N}_0}\{k'' \ | \ \exists (k'', 0)\text{-core } H \wedge v\in V_H\}$.
}
\label{def:Anchoredcoreness}
\end{definition}

Different from the undirected coreness, the anchored coreness is a two-dimensional feature of in-degree and out-degree in directed graphs. 
For example, consider the graph $G$ in Figure~\ref{fig:introexample} and $k=3$, the anchored coreness of vertex $v_2$ is $(3, 1)$, as $l_{max}(v_2, 3)=1$. Correspondingly, $\Phi(v_2)=\{(0, 2), (1, 2), (2, 2), (3, 1)\}$.
The anchored corenesses can facilitate the distributed D-core decomposition as follows. According to Property~\ref{pro.nest}, for a vertex $v$ with the anchored coreness of $(k, l)$, $v$ belongs to any $(k, l')$-core with $l' \leq l$. Hence, as long as we compute the anchored corenesses of $v$ for each possible $k$, we can get all D-cores containing $v$. As a result, for a given directed graph $G$, the problem of D-core decomposition is equivalent to computing the entire anchored corenesses for every vertex $v\in V_G$, i.e., $\{\Phi(v)| v\in V_G\}$.



\vspace*{-.2cm}
\subsection{Distributed Anchored Coreness Computing}
In this section, we present a distributed algorithm for computing the entire anchored corenesses for every vertex in $G$.


\stitle{Overview}. 
To handle the anchored coreness computation simultaneously in a distributed setting, we propose a distributed  vertex-centric algorithm to compute all feasible anchored corenesses $(k, l)$'s  for a vertex $v$.
The general idea is to first identify the feasible range of $k \in [0, k_{max}(v)]$ by exploring $(k, 0)$-cores and then refine an estimated upper bound of $l_{max}(v, k)$ to be exact for all possible values of  $k$.
The framework is outlined in Algorithm~\ref{algorithm:ACC}, which gives an overview of the anchored coreness updating procedure in three phases: 1) deriving $k_{max}(v)$; 2) computing the upper bound of $l_{max}(v, k)$ for each $k$; and 3) refining the upper bound to the exact anchored coreness $l_{max}(v, k)$.
Note that in the second and third phases, the upper bound of $l_{max}(v, k)$ can be computed and refined in batch, instead of one by one sequentially, for different values of $k \in [0, k_{max}(v)]$.

\begin{table*}[t]
\small
\vspace{-0.3cm}
\caption{An illustration of distributed D-core decomposition using Algorithm~\ref{algorithm:ACC} on graph $G$ in Figure~\ref{fig:exampleDcore}. }
\label{tab:exampleofacc}
\vspace{-0.4cm}
\begin{tabular}{cc|c|c|c|c|c|c|c|c|}
\cline{3-10}
\multirow{2}{*}{}                                & \multirow{2}{*}{} & \multicolumn{8}{c|}{Vertices}                                              \\ \cline{3-10}
                                                 &                   & $v_1$      & $v_2$      & $v_3$      & $v_4$      & $v_5$      & $v_6$      & $v_7$   & $v_8$      \\ \hline
\multicolumn{1}{|c|}{\multirow{3}{*}{Phase I}}   & $\mathrm{iH}^{(0)}(v)$          & 3       & 2       & 2       & 2       & 2       & 3       & 1    & 2       \\ \cline{2-10}
\multicolumn{1}{|c|}{}                           & $\mathrm{iH}^{(1)}(v)$          & \textbf{2}       & 2       & 2       & 2       & 2       & \textbf{2}       & 1    & 2       \\ \cline{2-10}
\multicolumn{1}{|c|}{}                           & $\mathrm{iH}^{(2)}(v)$  $=$ $k_{max}(v)$         & 2       & 2       & 2       & 2       & 2       & 2       & 1    & 2       \\ \hline
\multicolumn{1}{|c|}{\multirow{3}{*}{Phase II}}  & $\forall k \in [0, k_{max}(v)]$, $\mathrm{oH}_{G[k]}^{(0)}(v)$                 & 3; 3; 3 & 0; 0; 0 & 0; 0; 0 & 5; 5; 5 & 3; 3; 3 & 2; 2; 2 & 2; 2 & 2; 2; 2 \\ \cline{2-10}
\multicolumn{1}{|c|}{}                           & $\forall k \in [0, k_{max}(v)]$, $\mathrm{oH}_{G[k]}^{(1)}(v)$                 &  \textbf{2; 2; 2}       &0; 0; 0         & 0; 0; 0        &  \textbf{2; 2; 2}       &   \textbf{2; 2; 2}      &   2; 2; 2      &  2; 2    &  \textbf{1; 1; 0}       \\ \cline{2-10}
\multicolumn{1}{|c|}{}                           & $\forall k \in [0, k_{max}(v)]$, $\mathrm{oH}_{G[k]}^{(2)}(v)$                 &   2; 2; 2       &0; 0; 0         & 0; 0; 0        &  2; 2; 2       &   2; 2; 2      &   2; 2; 2      &  2; 2    &  1; 1; 0      \\ \hline
\multicolumn{1}{|c|}{\multirow{3}{*}{Phase III}} & $\forall k \in [0, k_{max}(v)]$, $l_{upp}(k, v)$                  &    2; 2; 2       &0; 0; 0         & 0; 0; 0        &  2; 2; 2       &   2; 2; 2      &   2; 2; 2      &  2; 2    &  1; 1; 0    \\ \cline{2-10}
\multicolumn{1}{|c|}{}                           & $\forall k \in [0, k_{max}(v)]$, $l'_{upp}(k, v)$                  &    2; 2; 2       &0; 0; 0         & 0; 0; 0        &  2; 2; 2       &   2; 2; 2      &   2; 2; 2      &  \textbf{2; 1}    &  1; 1; 0      \\ \cline{2-10}
\multicolumn{1}{|c|}{}                           & $\forall k \in [0, k_{max}(v)]$, $l_{max}(k, v)$                  &    2; 2; 2       &0; 0; 0         & 0; 0; 0        &  2; 2; 2       &   2; 2; 2      &   2; 2; 2      &  2; 1    &  1; 1; 0        \\ \hline
\end{tabular}
\end{table*}

\begin{algorithm}[t]
\linespread{0.8}\selectfont
\LinesNumbered
\caption{Distributed Anchored Coreness Computation: routine executed by a vertex $v$}
\label{algorithm:ACC}
\KwIn{directed graph $G$, vertex $v$}
\KwOut{anchored corenesses of vertex $v$}
  Compute $k_{max}(v)$ for vertex $v$ using Algorithm~\ref{algorithm:kmax};\\
  Compute the upper bounds $l_{upp}(k, v)$ where $k\in [0, k_{max}]$, by invoking Algorithm~\ref{algorithm:lUPC};\\
  Refine the upper bounds $l_{upp}(k, v)$ to  anchored corenesses $l_{max}(k, v)$  using Algorithm~\ref{algorithm:refine};\\
  \textbf{return} the entire anchored corenesses of $v$ as $\Phi(v)$;
\end{algorithm}


%



\stitle{Phase I: Computing the in-degree limit $k_{max}(v)$.}
To compute $k_{max}(v)$, first,  we introduce a concept of  H-index~\cite{H-index}.
 Specifically, given a collection of integers $S$, the H-index of $S$ is a maximum integer $h$ such that $S$ has at least $h$ integer elements whose values are no less than $h$, denoted as $\mathcal{H}(S)$. For example, given $S=\{1, 2, 3, 3, 4, 6\}$, H-index $\mathcal{H}(S)=3$, as $S$ has at least 3 elements whose values are  no less than 3. Based on H-index, we give a new definition of \emph{$n$-order in-H-index} for directed graph.

\begin{definition}
{\bf ($n$-order in-H-index).}  Given a vertex $v$ in $G$, the $n$-order in-H-index of $v$, denoted by $\mathrm{iH}_G^{(n)}(v)$, is defined as
\begin{equation}\label{inHindex}
\mathrm{iH}_G^{(n)}(v)=\begin{cases}
 deg^{in}_G(v), & n=0 \\
 \mathcal{H}(I), & n > 0
 \end{cases}
\end{equation}
\noindent where the integer set $I=\{\mathrm{iH}_G^{(n-1)}(u) | u \in N^{in}_G(v)\}$.
\label{def:ninHindex}
\end{definition}

\begin{theorem} [Convergence]
\begin{equation}
 k_{max}(v) = \lim_{n \rightarrow \infty} \mathrm{iH}_G^{(n)}(v)
\end{equation}
\label{theorem:kmaxconvergence}
\end{theorem}

\vspace{-0.4cm}

\begin{proof}
\gcc{Due to space limitation, we give a proof sketch here. The detailed proof can be found in~\cite{Arxiv-D3core}.} First, we prove that $\mathrm{iH}_G^{(n)}(v)$ is non-increasing with the increase of order $n$. Thus, $\mathrm{iH}_G^{(n)}(v)$ finally converges to an integer when $n$ is big enough. Then, we prove  $k_{max}(v) \leq \mathrm{iH}_{G'}^{(\infty)}(v) \leq \mathrm{iH}_G^{(\infty)}(v)$, where $G' \subseteq G$ is a subgraph induced by the vertices $v'$ with $k_{max}(v')\geq k_{max}(v)$. Also, we know $k_{max}(v) \geq \mathrm{iH}_G^{(\infty )}(v)$ by definition. Hence, $k_{max}(v) = \mathrm{iH}_G^{(\infty)}(v)$.
\end{proof}

According to Theorem~\ref{theorem:kmaxconvergence},  $\mathrm{iH}_G^{(n)}(v)$ finally converges to  $k_{max}(v)$,
based on which we present a distributed algorithm as shown in Algorithm~\ref{algorithm:kmax} to compute $k_{max}(v)$.
Algorithm~\ref{algorithm:kmax} has an initialization step (lines 1-4), and two update procedures after receiving one message (lines 5-7) and all messages (lines 8-11). It first uses 0 to initialize the set $I$, which keeps the latest $n$-order in-H-indexes of $v$'s in-neighbors (lines 1-2). Then, the algorithm sets the $n$-order in-H-index of $v$ to its in-degree (line 3) and sends the message <$v$, $\mathrm{iH}(v)$> to all its out-neighbors (line 4).
When $v$ receives a message <$v'$, $\mathrm{iH}(v')$> from its in-neighbor $v'$, the algorithm updates the $n$-order in-H-index of $v'$ (line 5). If  $\mathrm{iH}(v') < \mathrm{iH}(v)$, it means the $n$-order in-H-index of $v$ may decrease. Thus, \emph{flag} is set to \emph{True} to indicate the re-computation of  $v$'s  $n$-order in-H-index (line 7). After receiving all massages, if \emph{flag} is \emph{True}, Algorithm~\ref{algorithm:kmax} updates $v$'s  $n$-order in-H-index $\mathrm{iH}(v)$ and inform all its out-neighbors if $\mathrm{iH}(v)$ decreases (lines 9-11).  Algorithm~\ref{algorithm:kmax} completes and returns $\mathrm{iH}(v)$ as $k_{max}(v)$ when there is no vertex broadcasting messages (line 12).

\begin{example}
\label{example:kmax}
We use the directed graph $G$ in  Figure~\ref{fig:exampleDcore} to illustrate Algorithm~\ref{algorithm:kmax}, whose calculation process is shown in Table~\ref{tab:exampleofacc}.  We take vertex $v_1$ as an example. First, $v_1$'s 0-order in-H-index is initialized with its in-degree, i.e., $\mathrm{iH}_G^{(0)}(v_1)=3$. Then, Algorithm~\ref{algorithm:kmax} iteratively computes $\mathrm{iH}_G^{(n)}(v_1)$. After one iteration, the 1-order in-H-index of $v_1$ has converged to $\mathcal{S}(\mathrm{iH}_G^{(0)}(v_4), \mathrm{iH}_G^{(0)}(v_6), \mathrm{iH}_G^{(0)}(v_7))$ = $\mathcal{S}(2, 3, 1)$ = 2. Thus, $k_{max}(v_1) = \mathrm{iH}_G^{(2)}(v_1) = \mathrm{iH}_G^{(1)}(v_1) =2$.
\end{example}

\makeatletter
\newcommand{\nosemic}{\renewcommand{\@endalgocfline}{\relax}}
\newcommand{\dosemic}{\renewcommand{\@endalgocfline}{\algocf@endline}}
\newcommand{\pushline}{\Indp}
\newcommand{\popline}{\Indm\dosemic}
\let\oldnl\nl
\newcommand{\nonl}{\renewcommand{\nl}{\let\nl\oldnl}}
\makeatother
\begin{algorithm}[t]
\linespread{0.6}\selectfont
\caption{Computing $k_{max}(v)$}
\LinesNumbered
\label{algorithm:kmax}
\KwIn{directed graph $G$, vertex $v$}
\KwOut{$k_{max}(v)$}
\vspace{0.15cm}
\pushline\dosemic\nonl    \emph{\textbf{Initializations}}
\popline \\   \For{each $v' \in N^{in}_G(v)$}{
     $I[v']\gets 0$\;
  }
  $\mathrm{iH}(v)\gets deg^{in}_G(v)$\;
\nosemic  Send  message $\langle v$, $\mathrm{iH}(v) \rangle$ to all out-neighbors of $v$;\\
  \vspace{0.15cm}

 \pushline\dosemic\nonl\emph{\textbf{On receiving message} $\langle v'$, $\mathrm{iH}(v') \rangle$ \textbf{from} $v$'s \textbf{in-neighbor} $v'$}
\popline \\ $I[v'] \gets \mathrm{iH}(v')$\;
     \If{$\mathrm{iH}(v')$ < $\mathrm{iH}(v)$}{
\nosemic      $flag \gets$ True\;
     }

  \vspace{0.15cm}
\pushline\dosemic\nonl  \emph{\textbf{After receiving all  messages}}
\popline \\  \If{$flag=True$}{
     \If {$\mathcal{H}(I)$ < $\mathrm{iH}(v)$ }{
           $\mathrm{iH}(v) \gets \mathcal{H}(I)$;$flag \gets$ False\;
\nosemic           Send message $\langle v$, $\mathrm{iH}(v)\rangle$ to all out-neighbors of $v$\;

      }
   }
   \vspace{0.15cm}
\pushline\dosemic\nonl   \emph{\textbf{When no vertex broadcasts messages}}
\popline   \\ \Return{ $k_{max}(v) \gets \mathrm{iH}(v)$\;}
\end{algorithm}

\stitle{Phase II: Computing the upper bounds of \bm{$l_{max}(k, v)$}.}
In a distributed setting, the computation of $l_{max}(k, v)$ faces technical challenges.
It is difficult to compute $l_{max}(k, v)$ by making use of  only the ``intermediate'' neighborhood information. Because some vertices $u\in N_G(v)$ may become disqualified and thus be removed from the candidate set of $(k, l_{max}(k, v))$-core during the iteration process.
Even worse, verifying the candidacy of $u$ requires a large number of message exchanges between vertices. 
To address these issues, we design a novel upper bound for $l_{max}(k, v)$, denoted by $l_{upp}(k, v)$, which can be iteratively computed with ``intermediate'' corenesses to reduce communication costs.
To start with, we give a new definition of $n$-order out-H-index, similar to Definition~\ref{def:ninHindex}.

\begin{definition}
{\bf ($n$-order out-H-index).}  Given a vertex $v$ in $G$, the $n$-order out-H-index  of $v$, denoted as $\mathrm{oH}_G^{(n)}(v)$, is defined as
\begin{equation}\label{outHindex}
\mathrm{oH}_G^{(n)}(v)=\begin{cases}
 deg^{out}_G(v), & n=0 \\
 \mathcal{H}(O), & n > 0
 \end{cases}
\end{equation}
\noindent where $O=\{\mathrm{oH}_G^{(n-1)}(u) | u \in N^{out}_G(v)\}$.
\label{def:noutHindex}
\end{definition}

Based on $\mathrm{oH}_G^{(n)}(v)$, we have the following theorem.

\begin{theorem}
Given a vertex $v$ in $G$  and an integer $k\in [0, k_{max}(v)]$, let $G[k]$ be the subgraph of $G$ induced by the vertices in $V_k=\{u \ | \ u\in V_G \wedge k_{max}(u) \geq k\}$.
Then, it holds that
\begin{equation}
 l_{max}(k, v) \leq \lim_{n \rightarrow \infty} \mathrm{oH}_{G[k]}^{(n)}(v).
\end{equation}
\label{theorem:lmaxconvergence}
\end{theorem}

\vspace{-0.4cm}

\begin{proof}

%
\xkliao{
Similar to Theorem~\ref{theorem:kmaxconvergence}, we can prove $\lim \limits_{n \rightarrow \infty} \mathrm{oH}_{G[k]}^{(n)}(v)$ $ = l'$ such that $v\in (0,l')$-core of $G[k]$  but $v\notin (0, l'+1)$-core of $G[k]$. Then, we have the following relationship for the D-cores of $G[k]$: $(k, l_{max}(k, v))$-core $\subseteq$ $(0, l_{max}(k, v))$-core $\subseteq$  $(0,l')$-core. According to the partial nesting property of D-core, $l' \geq l_{max}(k, v)$ holds.
}
\end{proof}

Theorem~\ref{theorem:lmaxconvergence} indicates that  $\lim \limits_{n \rightarrow \infty} \mathrm{oH}_{G[k]}^{(n)}(v)$ can be served as an upper bound of $ l_{max}(k, v) $, i.e., $l_{upp}(k, v)= \lim \limits_{n \rightarrow \infty} \mathrm{oH}_{G[k]}^{(n)}(v)$. 
Thus, we can compute $l_{upp}(k, v)$ by iteratively calculating the $n$-order out-H-index of $v$ in the directed subgraph $G[k]$. Moreover, to efficiently compute $l_{upp}(k, v)$ for all values $k\in [0, k_{max}(v)]$  in parallel, our distributed algorithm should send updating messages in batch  and compute $l_{upp}(k, v)$ simultaneously.


\begin{algorithm}[t]
\linespread{0.8}\selectfont
\LinesNumbered
\caption{Computing Upper Bounds $l_{upp}(k, v)$}
\label{algorithm:lUPC}
\KwIn{directed graph $G$, vertex $v$, $k_{max}(v)$}
\KwOut{the upper bounds $l_{upp}(k, v)$ for $k \in [0, k_{max}(v)]$}
\vspace{0.1cm}
\pushline\dosemic\nonl   \emph{\textbf{Initializations}}
\popline \\ \For{each $k \in [0, k_{max}(v)]$}{
   \For{each $ v' \in N^{out}_{G[k]}(v)$}{
        $I[k][v'] \gets 0$\;
   }
   $\mathrm{oH}_v[k] \gets deg^{out}_{G[k]}(v)$; $change[k]\gets$ True\;
}

\nosemic Send  message $\langle v$, $\mathrm{oH}_v[\cdot], change[\cdot] \rangle$ to all in-neighbors of $v$;\\
  \vspace{0.1cm}

\pushline\dosemic\nonl  \emph{\textbf{On receiving message} $\langle v'$, $\mathrm{oH}_{v'}[\cdot], change[\cdot] \rangle$ \textbf{from} $v$'s \textbf{out-neighbor} $v'$}
\popline  \\ \For{each $k \in [0, k_{max}(v)]$}{
     \If{$change[k] =True$}{
        $I[k][v'] \gets \mathrm{oH}[k]$;\\
        \If{$\mathrm{oH}_{v'}[k]$ < $\mathrm{oH}_v[k]$}{
\nosemic           $flag[k] \gets$ True\;
        }
     }
  }

  \vspace{0.1cm}
\pushline\dosemic\nonl   \emph{\textbf{After receiving all  messages}}
\popline   \\ \For{each $ k \in [0, k_{max}(v)]$}{
   \If{$flag[k]=True$}{
     \If {$\mathcal{H}(I[k])$ < $\mathrm{oH}_v[k]$ }{
           $\mathrm{oH}_v[k] \gets \mathcal{H}(I[k])$; $change[k] \gets$ False\;

      }
   }
   }
    \If{$\exists k\in [0, k_{max}(v)]$, $change[k]=True$}{
\nosemic         Send  message $\langle v$, $\mathrm{oH}_v[\cdot], change[\cdot] \rangle$ to all in-neighbors of $v$;\\
   }

   \vspace{0.1cm}
\pushline\dosemic\nonl   \emph{\textbf{When no vertex broadcasts messages}}
\popline   \\ $l_{upp}[\cdot] \gets \mathrm{oH}_v[\cdot]$\;

\end{algorithm}

Based on the above discussion, we propose a distributed algorithm for computing the upper bounds $l_{upp}(k, v)$.
Algorithm~\ref{algorithm:lUPC} presents the detailed procedure. First, it initializes the $n$-order out-H-index of $v$ for each possible value of $k$ and sends them to $v$'s in-neighbors (lines 1-5). When $v$ receives a message from its out-neighbor $v'$, $v$ updates the $n$-order out-H-index of $v'$ for subsequent calculation (lines 6-10). After receiving all messages, $v$ updates its own $n$-order out-H-index for each possible value of $k$ (lines 11-14). If any $n$-order out-H-indexes of $v$ decreases, $v$ informs all its in-neighbors (lines 15-16). Finally, when there is no vertex broadcasting messages, we get the upper bound $l_{upp}(k, v)$ for each  $k \in [0, k_{max}(v)]$ (line 17).

\begin{example}
\label{example:lupp}
We illustrate Algorithm~\ref{algorithm:lUPC} by continuing Example~\ref{example:kmax}. As shown in Table~\ref{tab:exampleofacc}, since $k_{max}(v_1) = 2$, we first initialize the $0$-order out-H-indexes of $v_1$ as $\mathrm{oH}_{G[k]}^{(0)}(v_1)=3$  for each $k \in 
\{0, 1, 2\}$. 
After one iteration of computing the $n$-order out-H-indexes, all $1$-order out-H-indexes of $v_1$ 
have converged to 2. Thus, we have $\mathrm{oH}_{G[0]}^{(1)}(v_1) =2$, $\mathrm{oH}_{G[1]}^{(1)}(v_1) =2$, $\mathrm{oH}_{G[2]}^{(1)}(v_1) =2$.
\end{example}

\stitle{Phase III: Refining  \bm{$l_{upp}(k, v)$} to \bm{$l_{max}(k, v)$}.}
Finally, we present the third phase of refining the upper bound $l_{upp}(k, v)$ to get the exact anchored coreness $l_{max}(k, v)$. To this end, we first present the following theorem.

\begin{theorem}
Given a vertex $v$ in  $G$ and an integer $k$, if $(k, l_{upp}(k, v))$ is an anchored coreness of $v$, it should satisfy two constraints on in-neighbors and out-neighbors: \textbf{(i)} $v$ has at least $k$ in-neighbors $v'$ such that $l_{upp}(k, v') \geq l_{upp}(k, v)$; and \textbf{(ii)} $v$ has at least $l_{upp}(k, v)$ out-neighbors $v''$ such that $l_{upp}(k, v'') \geq l_{upp}(k, v)$.
\label{theorem:refine}
\end{theorem}

Theorem~\ref{theorem:refine} obviously holds, according to  Def.~\ref{def:dcore} of D-core and the upper bound $l_{upp}(k, v) \geq l_{max}(k, v)$.
Based on Theorem~\ref{theorem:refine}, we can refine $l_{upp}(k, v)$ decrementally by checking the upper bounds $l_{upp}(k, v')$'s of $v$'s in- and out-neighbors. If $v$ satisfies the above two constraints in Theorem~\ref{theorem:refine}, $l_{upp}(k, v)$ keeps unchanged; otherwise, $l_{upp}(k, v)$ decreases by 1 as the current $(k, l_{upp}(k, v))$ is not an anchored coreness of $v$. The above process needs to repeat for all vertices and all possible values of $k$, until none of $(k, l_{upp}(k, v))$ changes. Finally, we obtain all  anchored corenesses $\{\Phi(v)| v\in V_G\}$.

Algorithm~\ref{algorithm:refine} outlines the procedure of the distributed refinement phase. First, the algorithm initializes some auxiliary structures and  broadcast $v$'s upper bound  $l_{upp}(k, v)$ for each possible $k \in [0, k_{max}(v)]$ (lines 1-3).  When it receives a message from $v$'s neighbor $v'$, the algorithm updates the upper bound set for $v'$ (lines 4-7). After receiving all messages, the algorithm refines $l_{upp}(k, v)$ for each $k \in [0, k_{max}(v)]$ based on Theorem~\ref{theorem:refine} (lines 8-13). If there exists such a $(k, l_{upp}(k, v))$ whose $l_{upp}(k, v)$ is decreased, the algorithm broadcasts the new upper bound set to $v$'s neighbors (lines 14-15). As soon as there are no vertex broadcasting messages, Algorithm~\ref{algorithm:refine} terminates and we get all anchored corenesses of $v$ (lines 16-17).

\begin{example}
Continue Example~\ref{example:lupp} to illustrate Algorithm~\ref{algorithm:refine} in Phase III, which refines the upper bound $l_{upp}(k, v_1)$ to the exact $l_{max}(k, v_1)$. For  $k_{max}(v_1)=3$ and  each $ k\in [0, k_{max}(v_1)]$, Table~\ref{tab:exampleofacc} reports the final results $l_{upp}(k, v_1)$ $=$ $l_{max}(k, v_1) =2$. Therefore, the entire anchored corenesses of $v_1$ are $\Phi(v_1) = \{(0, 2), (1, 2), (2, 2)\}$.

\end{example}

\begin{algorithm}[t]
\linespread{0.8}\selectfont
\caption{Anchored Coreness Refinement}
\label{algorithm:refine}
\LinesNumbered
\KwIn{graph $G$, vertex $v$,  $k_{max}(v)$, upper bounds $l_{upp}[\cdot]$}
\KwOut{ the entire anchored corenesses of $v$ as $\Phi(v)$ }
\vspace{0.1cm}
\pushline\dosemic\nonl   \emph{\textbf{Initializations}}
\popline \\  \For{each $ k \in [0, k_{max}(v)]$}{
$change[k] \gets$ True; $l[k][v] \gets  0$;\\
}
\nosemic Send  message $\langle v$, $l_{upp}[\cdot], change[\cdot] \rangle$ to all neighbors of $v$;\\

\vspace{0.1cm}
\pushline\dosemic\nonl \emph{\textbf{On receiving message} $\langle v'$, $l_{upp}[\cdot], change[\cdot] \rangle$ \textbf{from} $v$'s \textbf{neighbor} $v'$}
\popline \\ \For{each $ k \in [0, k_{max}(v)]$}{
     \If{$change[k] =True$}{
        $l[k][v'] \gets l_{upp}[k]$;\\
\nosemic        $flag[k] \gets$ True\;
        }
     }

  \vspace{0.1cm}
\pushline\dosemic\nonl   \emph{\textbf{After receiving all  messages}}
\popline \\   \For{each $k \in [0, k_{max}(v)]$}{
   \If{$flag[k]=True$}{
     $V' \leftarrow\{v'|v'\in N^{in}_G(v) \wedge l[k][v'] \geq l_{upp}[k]\}$\;
     $V'' \leftarrow\{v''|v''\in N^{out}_G(v) \wedge l[k][v''] \geq l_{upp}[k]\}$\;
     \If {$|V'| < k$ \textbf{or}  $|V''| < l_{upp}[k]$ }{
           $l_{upp}[k] \gets l_{upp}[k]-1$;
           $change[k] \gets$ True\;

      }
   }
   }

    \If{$\exists$$k\in [0, k_{max}(v)]$ such that $change[k]=True$ }{
\nosemic         Send  message $\langle v$, $l_{upp}[\cdot], change[\cdot] \rangle$ to all neighbors of $v$;\\
   }

   \vspace{0.1cm}
\pushline\dosemic\nonl   \emph{\textbf{When no vertex broadcasts messages}}
\popline \\   \For{each $k \in [0, k_{max}(v)]$}{
      Add $(k, l_{upp}[k])$ to the anchored corenesses $\Phi(v)$\;
   }
\end{algorithm}

\subsection{Algorithm Analysis and Extension}

\stitle{Complexity analysis}. We first analyze the time, space, message complexities of Algorithm~\ref{algorithm:ACC}. Let the edge size $|E_G| = m$,  the maximum in-degree $\Delta_{in}= \max_{v\in V_G} deg^{in}_G(v)$, the maximum out-degree $\Delta_{out}= \max_{v\in V_G} deg^{out}_G(v)$, and  the maximum degree $\Delta= \max_{v\in V_G} deg_G(v)$. \gcc{
In addition, let $R_{AC-I}$, $R_{AC-II}$, and $R_{AC-III}$ be the number of  convergence rounds required by  the three phases in Algorithm~\ref{algorithm:ACC}, respectively. Let be the total number of converge rounds in Algorithm~1 as $R_{AC} = R_{AC-I}+R_{AC-II}+R_{AC-III}$ and $R_{AC} \in O(\Delta)$.
} We have the following theorems (their detailed proofs can be found in ~\cite{Arxiv-D3core}): 

\begin{theorem}(\textbf{Time and Space Complexities})
\gcc{Algorithm~\ref{algorithm:ACC} takes $O(R_{AC} \cdot \Delta_{in}  \cdot \Delta)$ time and $O(\Delta_{in}  \cdot \Delta)$ space. The total time and space complexities for computing all vertices' corenesses are $O(R_{AC} \cdot \Delta_{in} \cdot m)$ and $O(\Delta_{in} \cdot m )$, respectively.}
\label{theorem:ACTC}
\end{theorem}

\begin{theorem}(\textbf{Message Complexity})
The message complexity (i.e., the total number of times that a vertex send messages) of Algorithm~\ref{algorithm:ACC} is $O(\Delta_{in}  \cdot \Delta_{out} \cdot \Delta)$. The total message complexity for computing all vertices' corenesses is  $O(\Delta_{in}  \cdot \Delta_{out} \cdot  m)$.
\label{theorem:ACMC}
\end{theorem}

\vspace*{-.2cm}
\stitle{Block-centric extension of Algorithm~\ref{algorithm:ACC}}. We further discuss to extend the vertex-centric D-core decomposition in Algorithm~\ref{algorithm:ACC} to the block-centric framework.
The extension can be easily achieved by changing the update operation after receiving all messages. That is,
instead of having Algorithms~\ref{algorithm:kmax}, \ref{algorithm:lUPC} and \ref{algorithm:refine}  perform the update operation only once after receiving all messages,  in the block-centric framework, the algorithms should update the H-indexes multiple times until the local block converges. For example, Algorithm~\ref{algorithm:kmax} computes the $n$-order in-H-index of $v$ only once in each round (lines 10-13). In contrast, the block-centric version should compute $v$'s $n$-order in-H-index iteratively with $v$'s in-neighbors, that are located in the same block as $v$, before broadcasting messages to other blocks to enter the next round. \gcc{
Note that in the worst case, for block-centric algorithms, every vertex converges within the block after computing the in-H-index/out-H-index only once, which is the same as vertex-centric algorithms. Therefore, the worse-case cost of block-centric algorithms is the same as that of vertex-centric algorithms.}

\section{Distributed Skyline Coreness-Based Algorithm}
\label{Sec:SCC}
In this section, we propose a novel concept of skyline coreness, which is more elegant than the anchored coreness. Then, we give a new definition of $n$-order D-index for computing skyline corenesses. Based on the D-index, we propose a distributed algorithm for skyline coreness computation to accomplish D-core decomposition.

\vspace*{-.3cm}
\subsection{Skyline Coreness}
\stitle{Motivation}. The motivation for proposing another skyline coreness lies in an important observation that the anchored corenesses $(k, l)$'s  may have redundancy.
For example, in Figure~\ref{fig:introexample}, the vertex $v_2$ has four anchored corenesses, i.e., $\Phi(v_2) = \{ $(0, 2), (1, 2), (2, 2),  (3, 1)$ \}$. According to D-core's partial nesting property, if $v_2 \in (2, 2)$-core, $v_2$ must also belong to $(0, 2)$-core and $(1, 2)$-core. Thus, it is sufficient and more efficient to keep the coreness of $v_2$ as $\{$(2, 2), (3, 1)$\}$, which uses (2, 2)-core to represent other two D-cores (0, 2)-core and (1, 2)-core. This elegant representation is termed as \emph{skyline coreness},  which can facilitate space saving and fast computation of D-core decomposition. Based on the above observation, we formally define the dominance operation and skyline coreness as follows.

\begin{definition}
{\bf (Dominance Operations).} Given two coreness pairs $(k, l)$ and $(k', l')$, we define
two operations `$\prec$' and `$\preceq$'  to compare them:
(i) $(k', l') \prec (k, l)$ indicates that $(k, l)$ dominates $(k', l')$, i.e., either $k'<k$, $l' \leq l$ hold or $k'\leq k$, $l' < l$ hold; and (ii) $(k', l') \preceq (k, l)$ represents that $k' \leq  k$, $l' \leq l$  hold.
\label{def:dominance}
\end{definition}

\begin{definition}
{\bf (Skyline Coreness).} Given a vertex $v$ in a directed graph $G$ and a coreness pair $(k, l)$, we say that $(k, l)$ is a skyline coreness of $v$ \emph{iff} it satisfies that (i) $v \in (k, l)$-core; and (ii) there exist no other pair $(k', l')$ such that  $(k, l) \prec (k', l')$ and $v \in (k', l')$-core. We use $\SC(v)$ to denote the \emph{entire skyline corenesses} of the vertex $v$, i.e., $\SC(v)=\{(k, l) \ |\  (k, l)$ is a skyline coreness of $v\}$.
\label{def:skylineness}
\end{definition}

In other words, the skyline coreness of a vertex $v$ is a non-dominated pair $(k, l)$ whose corresponding $(k, l)$-core contains $v$.  For instance, vertex $v_2$ has the skyline corenesses $\SC(v_2) =\{(2, 2), (3, 1)\}$ in Figure~\ref{fig:introexample}, reflecting that  no other coreness $(k, l) $ can dominate any skyline coreness in $\SC(v_2)$.
According to D-core's partial nesting property, for a skyline coreness $(k, l)$ of $v$, $v$ is contained in the $(k', l')$-core with $(k', l') \prec (k, l)$. Therefore, if we compute all skyline corenesses $\SC(v)$ for a vertex $v$, we can find all D-cores the vertex $v$ belonging to. As a result, the problem of D-core decomposition is equivalent to computing the entire skyline corenesses for every vertex in $G$, i.e., $\{\SC(v)| v\in V_G\}$.

\stitle{Structural properties of skyline coreness}. We analyze the structural  properties of skyline coreness.

\begin{property}
Let $(k_v, l_v)$ be a skyline coreness of $v$, the following properties hold:
\begin{itemize}
\item[(I)] There exist $k_v$ in-neighbors $v' \in N_G^{in}(v)$ such that $(k_v, l_v)$ $\preceq$ $(k_{v'}, l_{v'})$, and also  $l_v$ out-neighbors $v''\in N_G^{out}(v)$ such that $(k_v, l_v)$ $\preceq$ $(k_{v''}, l_{v''})$.
\item[(II)] Two cases \textbf{cannot} hold in either way: there exist $k_v+1$ in-neighbors $v'\in N_G^{in}(v)$ such that $(k_v+1, l_v)$ $\preceq$ $(k_{v'}, l_{v'})$, or $l_v$ out-neighbors $v''\in N_G^{out}(v)$ such that $(k_v+1, l_v)$ $\preceq$ $(k_{v''}, l_{v''})$.
\item[(III)] Two cases \textbf{cannot} hold in either way: there exist  $k_v$ in-neighbors $v'\in N_G^{in}(v)$ such that $(k_v, l_v+1)$ $\preceq$ $(k_{v'}, l_{v'})$, or  $l_v+1$ out-neighbors $v''\in N_G^{out}(v)$ such that $(k_v, l_v+1)$ $\preceq$ $(k_{v''}, l_{v''})$.
\end{itemize}




\label{property:skylinecoreness}
\end{property}

\begin{proof}
First, we prove Property~\ref{property:skylinecoreness}(I). Let $D_1$ be the $(k_v, l_v)$-core of $G$, we have $deg_{D_1}^{in}(v) \geq k_v$ and  $deg_{D_1}^{out}(v) \geq l_v$.  For $\forall v' \in (N_{D_1}^{in}(v) \cup N_{D_1}^{out}(v))$, $v'$ may  be in the $(k', l')$-core with $k_v  \leq k' \leq k_{v'}$ and $l_v  \leq l' \leq l_{v'}$. Therefore, (I) of Property~\ref{property:skylinecoreness} holds.

Next, we prove Property~\ref{property:skylinecoreness}(II). Assume that $v$ has  $k_v+1$ in-neighbors $V'$  and $l_v$ out-neighbors $V''$ satisfying the constraints of (II).  Then, $V'$ and $V''$ must be in the $(k_v+1, l_v)$-core. Moreover, $v$ $\cup$ $(k_v+1, l_v)$-core is also a  $(k_v+1, l_v)$-core. Hence, $(k_v+1, l_v)$ rather than $(k_v, l_v)$ is a skyline coreness of $v$, which contradicts to the condition of Property~\ref{property:skylinecoreness}. Therefore, the assumption does not hold.

Finally, Property~\ref{property:skylinecoreness}(III) can be proved in the same way of  Property~\ref{property:skylinecoreness}(II). It is omitted due to space limitation.
\end{proof}

For example, $(2, 2)$ is a skyline coreness of $v_2$ in Figure~\ref{fig:introexample}. The in-neighbors of $v_2$ are  $v_3$, $v_4$, $v_5$, and $v_7$, whose skyline corenesses are $\{(3, 3)\}$, $\{(2, 2)\}$, $\{(3, 3)\}$, and $\{(2, 2), (3, 1)\}$, respectively. These four vertices all have skyline corenesses that dominate or are identical to $v_2$'s skyline coreness $(2, 2)$.
But only two vertices $v_3$ and $v_5$ have skyline corenesses that dominate $(k_{v_2}+1, l_{v_2}) = (3,2)$. Hence, $(3,2)$ is not a skyline coreness of $v_2$. Property~\ref{property:skylinecoreness} reveals the  relationships among vertices' skyline corenesses, based on which we propose 
an algorithm for skyline coreness computation in the next subsection.


\subsection{Distributed Skyline Corenesses Computing}

We begin with a novel concept of D-index.





\begin{definition}
{\bf (D-index).}  Given two sets of pairs of integers $R_{in}$, $R_{out}$ $\subseteq \mathbb{N}_0\times\mathbb{N}_0$, the D-index of $R_{in}$ and $R_{out}$ is denoted by $\mathcal{D}(R_{in}, R_{out})$ $\subseteq \mathbb{N}_0\times\mathbb{N}_0$, where each element
 $(k, l)\in \mathcal{D}(R_{in}, R_{out})$ satisfies:
(i) there exist at least $k$ pairs $(k_i, l_i) \in R_{in}$ such that $(k, l) \preceq (k_i, l_i)$ for $1\leq i\leq k$;
(ii) there exist at least $l$ pairs $(k_j, l_j) \in R_{out}$ such that $(k, l) \preceq (k_j, l_j)$ for $1\leq j\leq l$;
(iii) there does not exist another $(k', l') \in \mathbb{N}_0\times\mathbb{N}_0$ satisfying the above conditions (1) and (2), and $(k, l) \prec (k', l')$.



\label{def:Dindex}
\end{definition}

The idea of D-index is very similar to H-index.  Actually, the D-index is an extension of H-index to handle two-dimensional integer pairs. For $\mathcal{D}(R_{in}, R_{out})$, it finds a series of $(k, l)$ skyline pairs such that each has at least $k$ dominated pairs  in  $R_{in}$ and at least $l$ dominated pairs in $R_{out}$, using a joint indexing way.
 For example, let $R_{in}=\{(1, 1), (2, 2)\}$ and $R_{out} = \{(3, 3), (4, 4)\}$, then $\mathcal{D}(R_{in}, R_{out})=\{(1, 2)\}$. Note that  $\mathcal{D}(R_{in}, R_{out}) \neq \mathcal{D}(R_{out}, R_{in})$ may hold for the D-index, as  $\mathcal{D}(R_{out}, R_{in}) = \{(2, 1)\} \neq \{(1, 2)\} = \mathcal{D}(R_{in}, R_{out})$ in this example. 
 Next, we introduce another concept of $n$-order D-index for distributed D-core decomposition.


\begin{definition}
\label{def:norderDindex}
{\bf ($n$-order D-index).}  Given a vertex $v$ in  $G$, the $n$-order D-index  of $v$, denoted by $D^{(n)}(v) $ $\subseteq \mathbb{N}_0\times\mathbb{N}_0$, is defined as
\begin{equation}
D^{(n)}(v)=\begin{cases}
 \{(deg_G^{in}(v), deg_G^{out}(v))\}, & n=0 \\
\mathcal{D}(R_{in}^{(n-1)}(v), R_{out}^{(n-1)}(v)), & n > 0
 \end{cases}
\end{equation}

\noindent Here, $R_{in}^{(n-1)}(v)$ $=$ $\{(k_{u}, l_{u}) \in D^{(n-1)}(u) \ | \  u \in N_G^{in}(v)\}$ and $R_{out}^{(n-1)}(v)$ $=$ $\{(k_{u}, l_{u}) \in D^{(n-1)}(u) \ | \  u \in N_G^{out}(v)\}$.
Note that $D^{(n)}(v)$ is the \emph{largest  non-dominated} D-index such that it dominates or at least is identical to $\mathcal{D}(R_{in}^{(n-1)}(v), R_{out}^{(n-1)}(v))$, for each $R_{in}^{(n-1)}(v)$ $\in  D^{(n-1)}(u_1) \times \ldots \times  D^{(n-1)}(u_i)$ when $N_G^{in}(v) = \{u_1, \ldots, u_i\}$ and each $R_{out}^{(n-1)}(v)$ $\in  D^{(n-1)}(u_1) \times \ldots \times  D^{(n-1)}(u_j)$ when $N_G^{out}(v) = \{u_1, \ldots, u_j\}$.



\end{definition}

The $n$-order D-index $D^{(n)}(v)$ may contain more than one pair $(k, l)$,  i.e.,  $|D^{(n)}(v)| \geq 1$. Note that $R_{in}^{(n-1)}(v)$ and $R_{out}^{(n-1)}(v)$ consist of one pair $(k_u, l_u)$ for each in-neighbor $u\in N^{in}_G(v)$ and each out-neighbor $u\in N^{out}_G(v)$, respectively. Therefore, there exist multiple combinations of  $R_{out}^{(n-1)}(v)$ and $R_{out}^{(n-1)}(v)$. Moreover, $D^{(n)}(v)$
should consider all combinations of $R_{out}^{(n-1)}(v)$ and $R_{out}^{(n-1)}(v)$, and finally select the ``best'' choice as the \emph{largest non-dominated} set of D-index $\mathcal{D}(R_{in}^{(n-1)}(v), R_{out}^{(n-1)}(v))$.


For two pair sets $R_1$, $R_2$ $\subseteq \mathbb{N}_0\times\mathbb{N}_0$, we say $R_2 \preceq R_1$ if and only if $\forall (k, l)\in R_2$,  $\exists (k', l') \in R_1$ such that $(k, l)$ $\preceq $ $(k', l')$.
Then, we have the following theorem of $n$-order D-index convergence.

\begin{theorem} [$n$-order D-index Convergence]
For a vertex $v$ in $G$, it holds that
\begin{equation}
\SC(v) = \lim_{n \rightarrow \infty} D^{(n)}(v)
\end{equation}
\label{theorem:dindexconvergence}
\end{theorem}
\vspace{-0.5cm}
\begin{proof}
The proof can be similarly done as Theorem~\ref{theorem:kmaxconvergence}.
\end{proof}

By Theorem~\ref{theorem:dindexconvergence}, we can compute vertices' skyline corenesses via iteratively computing their $n$-order D-indexes until convergence.

\subsection{Algorithms and Optimizations}
A naive implementation of the distributed algorithm to compute $D^{(n)}(v)$ may suffer from serious performance problems, due to the combinatorial blow-ups in a large number of choices of $R_{in}^{(n-1)}(v)$ and $R_{out}^{(n-1)}(v)$. Thus, we first tackle three critical issues for fast distributed computation of $n$-order D-index.

\stitle{Optimization-1: Fast computation of D-index $\mathcal{D}(R_{in}, R_{out})$}. The first issue is, given $R_{in}$ and $R_{out}$, how to compute D-index $\mathcal{D}(R_{in}, R_{out})$. A straightforward way is to list all candidate pairs and return the pairs satisfying Def.~\ref{def:Dindex}. According to conditions (1)\&(2) in Def.~\ref{def:Dindex}, if $(k, l)$ belongs to D-index, there exists at least $k$ pairs of $R_{in}$ satisfying the dominance relationship. Therefore, $0 \leq k \leq |R_{in}|$. Similarly,  $0 \leq l \leq |R_{out}|$. Thus, there are a total of $(|R_1|+1) \cdot (|R_2|+1)$ candidate pairs to be checked, which is costly for large $|R_1|$ and $|R_2|$. In addition, the basic operation of D-index computation is frequently invoked in the process of computing $D^{(n)}(v)$. Hence, it is necessary to develop faster algorithms. To this end, we try to reduce the pairs for examination as many as possible through the following two optimizations.

\begin{itemize}
  \item \emph{Reducing the ranges of $k$ and $l$}. For conditions (1)\&(2) in Def.~\ref{def:Dindex}, if $(k, l)$ belongs to $\mathcal{D}(R_{in}, R_{out})$, there exist at least $k$   pairs  $(k_i, l_i)$ in $R_{in}$ such that $(k, l) \preceq (k_i, l_i)$. In other words, at least $k$   pairs  $(k_i, l_i)$ in $R_{in}$ have  $k_i \geq k$. Thus, the maximum $k$ is denoted by $k_{max}$ $=$ $\mathcal{H}(I_k)$, where $I_k=\{k_i \ | \ (k_i, l_i) \in R_{in}\}$. Similarly, we can also obtain the maximum $l$, denoted by $l_{max}$, as $l_{max}$ $=$ $\mathcal{H}(O_l)$, where $O_l=\{l_j \ | \  (k_j, l_j) \in R_{out}\}$. Since $\mathcal{H}(I_k) \leq |R_{in}|$ and $\mathcal{H}(O_l) \leq |R_{out}|$,  the total number of  candidate pairs decreases.
  \item  \emph{Pruning disqualified candidate pairs}.  Let $(k, l)$ $\in$ $\mathcal{D}(R_{in}, R_{out})$. According to condition (3) in Def.~\ref{def:Dindex}, if  $(k', l')$ $\in$ $\mathcal{D}(R_{in}, R_{out})$ with $k' <k$, $l'$ must satisfy $l' >l$. Otherwise, $(k', l') \prec (k, l)$ and  $(k', l')$ $\notin$ $\mathcal{D}(R_{in}, R_{out})$. This rule can be used to prune disqualified pairs based on the found skyline corenesses.
\end{itemize}

\stitle{Optimization-2: Fast computation of n-order D-index $D^{(n)}(v)$}.  The second issue is the computation of $D^{(n)}(v)$. By Def.~\ref{def:norderDindex}, both $R_{in}^{(n-1)}(v)$ and $R_{out}^{(n-1)}(v)$ may have multiple instances. Hence, a straightforward way is to compute the D-index for every instance and finally integrate them together. In total, we need to compute the D-index $O(\prod_{v' \in N_G^{in}(v)} |D^{(n-1)}(v')|$ $\cdot$ $\prod_{v'' \in N_G^{out}(v)} |D^{(n-1)}(v'')|)$ times, which is very inefficient. Actually, several redundant computations occur due to many independent instances in the D-index computation.
For example, in one instance, we have verified that $(k, l)$ belongs to the $n$-order D-index. Then, there is no need to verify $(k, l)$  in other instances. This motivates us to devise a more efficient method to compute $D^{(n)}(v)$, 
which requires D-index computation only once. Specifically, we first compute $k_{max}$ and $l_{max}$. Then, we enumerate candidate pairs for dominance checking. Here, we highlight two differences from the original D-index computation method.
\begin{itemize}
  \item \emph{The difference of $k_{max}$ and $l_{max}$ computations}.  For $k_{max}$ and $l_{max}$ in D-index computation, $I_k$ (resp. $O_l$) is formed by just adding $k_i$ (resp. $l_i$) from each pair in $R_{in}$. For $n$-order D-index computation, the vertex's $(n-1)$-order D-index may have more than one pairs. We should select the maximum $k_i$ and $l_i$ among these pairs. Specifically, for $v$'s $n$-order D-index computation, to compute $k_{max}$,  $I_k(v)=\{k_i \ | \ v' \in N_G^{in}(v), k_i= \max_{(k'_i, l'_i) \in D^{(n-1)}(v')}(k'_i)\}$. In the same way, $I_l(v)=\{l_j \ | \ v' \in N_G^{out}(v), l_j= \max_{(k'_j, l'_j) \in D^{(n-1)}(v')}(l'_j)\}$.

  \item \emph{The difference of dominance checking.} For a candidate pair $(k, l)$, the D-index computation should find the pairs in $R_{in}$ and $R_{out}$ that dominate or are identical to $(k, l)$. To compute $D^{(n)}(v)$, we should find all $v$'s neighbors $v'$ whose $(n-1)$-order D-index has a pair  dominating or identical to $(k, l)$. If $D^{(n-1)}(v')$  has multiple pairs, we need to examine the dominance relationship for each of these  pairs  with  $(k, l)$. Once one pair dominates or is identical to $(k, l)$, such $v'$ is identified.
\end{itemize}

\begin{algorithm}[t]
\linespread{0.8}\selectfont
\caption{Distributed Skyline Corenesses Computation  Algorithm: routine executed by vertex  $v$ }
\label{algorithm:skylinecoreness}
\LinesNumbered
\KwIn{directed graph $G$, vertex $v$}
\KwOut{the skyline corenesses $\SC(v)$ }
\vspace{0.1cm}
\pushline\dosemic\nonl   \emph{\textbf{Initializations}}
\popline \\  Compute $\mathrm{iH}^{(\infty)}_G(v)$ and $\mathrm{oH}^{(\infty)}_G(v)$ using Algorithm~\ref{algorithm:kmax};\\
$D_v=\{(\mathrm{iH}^{(\infty)}_G(v), \mathrm{oH}^{(\infty)}_G(v))\}$;\\
Send  message $\langle v$, $D_v \rangle$ to all neighbors of $v$;\\

\vspace{0.1cm}
\pushline\dosemic\nonl   \emph{\textbf{On receiving message$\langle v'$, $D_{v'}\rangle$ \textbf{from} $v$'s \textbf{neighbor} $v'$}}
\popline \\ $D_k[v']  \leftarrow 0$; $D_l[v']  \leftarrow 0$;\\
$D[v']  \leftarrow D_{v'}$;\\
\For{each $(k, l) \in D_{v'}$}{
    $D_k[v'] \leftarrow \max(D_k[v'], k)$; $D_l[v'] \leftarrow \max(D_l[v'], l)$;\\
}
$flag \gets$ True\;

\vspace{0.1cm}
\pushline\dosemic\nonl   \emph{\textbf{After receiving all  messages}}
\popline \\
\If{$flag=True$}{
Apply Algorithm~\ref{algorithm:Dindexcomputation} on $n$-order D-index computation;

\If{$D[v] \neq D$}{
        $D[v] \leftarrow D$; $D_v \leftarrow D$;\\
         Send  message $\langle v$, $D_v \rangle$ to all neighbors of $v$;\\
   }
}

\vspace{0.1cm}
\pushline\dosemic\nonl   \emph{\textbf{When no vertex broadcasts messages}}
\popline \\
\Return $\SC(v) \gets D[v]$;
\setlength{\floatsep}{0pt}
\end{algorithm}

\begin{algorithm}[t]
\linespread{0.8}\selectfont
\caption{$n$-order D-index Computation }
\label{algorithm:Dindexcomputation}
\LinesNumbered
\KwOut{$v$'s $n$-order D-index }
    $D \gets \varnothing$; $l_{min} \gets 0$;\\
    $I_k=\{D_k[v']| v' \in N_G^{in}(v)\}$; $k_{max}\leftarrow \mathcal{H}(I_k)$;\\
    $O_l=\{D_l[v']| v' \in N_G^{out}(v)\}$; $l_{max}\leftarrow \mathcal{H}(O_l)$;\\

    \For{$k \gets k_{max}$ \KwTo $0$}{
        $l \gets l_{max}$;\\
        \While{$l > l_{min}$}{
           $V_1 = \{v'|v' \in N_G^{in}(v),$ and $\exists (k', l') \in D[v'],$ $(k, l) \preceq (k', l')\}$;\\
           $V_2 = \{v'|v' \in N_G^{out}(v),$ and $\exists (k', l') \in D[v'],$ $(k, l) \preceq (k', l')\}$;\\
           \If{$|V_1| \geq k$ $\wedge$ $|V_2| \geq l$}{
               $l_{min} \leftarrow l$; $D \leftarrow D \cup (k, l)$; \\
           }
           $l \leftarrow l-1$ ;\\
        }
    }
\end{algorithm}

\begin{table*}[t]
\caption{An illustration of distributed skyline coreness computation using Algorithm~\ref{algorithm:skylinecoreness} on graph $G$ in Figure~\ref{fig:exampleDcore}. }
\small
\vspace{-0.2cm}
\label{tabshprocess}
\begin{tabular}{c|c|c|c|c|c|c|c|c|}
\cline{2-9}
\multirow{2}{*}{}       & \multicolumn{8}{c|}{\textbf{Vertices}}                         \\ \cline{2-9}
                        & $v_1$ & $v_2$ & $v_3$ & $v_4$ & $v_5$ & $v_6$ & $v_7$ & $v_8$ \\ \hline
\multicolumn{1}{|c|}{$D^{(0)}(v)$} &  $\{(2, 2)\}$    &  $\{(2, 0)\}$    &  $\{(2, 0)\}$    &  $\{(2, 2)\}$    & $\{(2, 2)\}$     &  $\{(2, 2)\}$    &  $\{(1, 2)\}$    & $\{(2, 1)\}$     \\ \hline
\multicolumn{1}{|c|}{$D^{(1)}(v)$} &  $\{(2, 2)\}$    &  $\{(2, 0)\}$    &  $\{(2, 0)\}$    &  $\{(2, 2)\}$    & $\{(2, 2)\}$     &  $\{(2, 2)\}$    &  $\{(0, 2), (1, 1)\}$    &  $\{(1, 1), (2, 0)\}$    \\ \hline
\multicolumn{1}{|c|}{$D^{(2)}(v)$} &  $\{(2, 2)\}$    &   $\{(2, 0)\}$   &  $\{(2, 0)\}$   &   $\{(2, 2)\}$   &  $\{(2, 2)\}$    &   $\{(2, 2)\}$   &  $\{(0, 2), (1, 1)\}$    &  $\{(1, 1), (2, 0)\}$    \\ \hline
\end{tabular}
\end{table*}

\stitle{Optimization-3: Tight initialization}. Finally, we present an optimization for $D^{(n)}(v)$ computation using a tight initialization. In Def.~\ref{def:norderDindex}, the $0$-order D-index is initialized with the vertex's in-degree and out-degree. The optimization idea is that if we tightly initialize the vertex's $0$-order D-index with smaller values (denoted by $D^{(0)}(v) = (k_0(v), l_0(v))$), the $n$-order D-index can converge faster to the exact skyline coreness. Here, we highlight two principles to find such $(k_0(v), l_0(v))$: (i) $k_0(v) \leq \max_{(k_i, l_i) \in \SC(v)} k_i$ and $l_0(v) \leq \max_{(k_i, l_i) \in \SC(v)} l_i$, otherwise the $D^{(n)}(v)$  cannot converge to $\SC(v)$; (ii) $(k_0(v), l_0(v))$ should be easy to compute in distributed settings. As a result, we present the following theorem.

\vspace*{-.3cm}
\begin{theorem}
For any vertex $v$ in $G$, it holds that $ k_{max}(v) \geq \max \{k_i \ | \ (k_i, l_i) \in \SC(v)\}$ 
 and $l_{max}(v) \geq \max \{l_i \ | \ (k_i, l_i) \in \SC(v)\}$, where $l_{max}(v)$ $=$ $\max \{l \ | \ v\in (0, l)\text{-core } \wedge v\notin (0, l+1)\text{-core}\}$.
\label{theorem:x0y0}
\end{theorem}

\vspace*{-.3cm}
Theorem~\ref{theorem:x0y0} offers two tight upper bounds for $k_0$ and $l_0$, i.e., $k_{max}(v)$ and $l_{max}(v)$, respectively. In addition, according to Theorems~\ref{theorem:kmaxconvergence} and~\ref{theorem:lmaxconvergence}, $k_{max}(v)$ and $l_{max}(v)$ can be computed by iteratively computing $v$'s $n$-order in-H-index and out-H-index, respectively. Therefore, we initialize $D^{(0)}(v)= (k_{max}(v), l_{max}(v))$.


\stitle{Algorithms}. Based on the above theoretical analytics and optimizations, we present the distributed skyline corenesses computation algorithm in Algorithm~\ref{algorithm:skylinecoreness}. At the initialization phase, the algorithm computes $\mathrm{iH}^{(\infty)}_G(v)$ and $\mathrm{oH}^{(\infty)}_G(v)$ using Algorithm~\ref{algorithm:kmax} and uses them to initialize the 0-order D-index of $v$, which is broadcast to all neighbors of $v$ (lines 1-3). When $v$ receives a message from its neighbor $v'$,  Algorithm~\ref{algorithm:skylinecoreness} updates the $n$-order D-index of $v'$ that is stored in $v$'s node, and finds the maximum values in each pair of $k$ and $l$ (lines 4-8). After $v$ receives all messages, Algorithm~\ref{algorithm:skylinecoreness} computes the $n$-order D-index for $v$, which is described in Algorithm~\ref{algorithm:Dindexcomputation}. Then, it broadcasts to all neighbors of $v$ if the $n$-order D-index changes (lines 9-13). When there is no vertex broadcasting messages, Algorithm~\ref{algorithm:skylinecoreness} returns the latest $n$-order D-index as skyline corenesses 
(line 14).

Next, we present the procedure of Algorithm~\ref{algorithm:Dindexcomputation} for $n$-order D-index computation. It first computes $k_{max}$ and $l_{max}$ as shown in Optimization-1 and~Optimization-2 (lines 2-3), which help to determine the range of candidate pairs. Then, the algorithm enumerates all candidate pairs $(k, l)$ and examines whether $(k, l)$ belongs to the $n$-order D-index of $v$  (lines 6-11). Note that $l_{min}$ keeps the minimal value of $l$ for the remaining candidate pairs, which is used to prune disqualified pairs.

\begin{example}
We use the graph $G$ in Figure~\ref{fig:exampleDcore} to illustrate Algorithm~\ref{algorithm:skylinecoreness}. Table~\ref{tabshprocess} reports the process of computing skyline corenesses. Take vertex $v_7$ as an example. First, the 0-order D-index of $v_7$ is initialized with $\{(1, 2)\}$, i.e., $D^{(0)}(v_7) =\{(1, 2)\}$. Then, we iteratively compute the $n$-order D-index for $v_7$. We can observe that after one iteration only, the $1$-order D-index of $v_7$ has converged as   $D^{(2)}(v_7) =D^{(1)}(v_7) =$ $\{(0, 2), (1, 1)\}$. Thus, the entire skyline corenesses of $v_7$ are $\SC(v_7)= $ $\{(0, 2), (1, 1)\}$.
\end{example}


\subsection{Algorithm Analysis and Extension}




\stitle{Complexity analysis.}
 \gcc{Let $R_{SC}$ be the number of convergence rounds 
taken by Algorithm~\ref{algorithm:skylinecoreness}. In practice, our algorithms achieve $R_{SC} \leq R_{AC} \ll \Delta$ on real datasets. We show the time, space, and message complexities of Algorithm~\ref{algorithm:skylinecoreness} below.} 

\vspace*{-.1cm}

\begin{theorem}(\textbf{Time and Space Complexities})
\gcc{Algorithm~\ref{algorithm:skylinecoreness} takes $O(R_{SC} \cdot \Delta_{in} \cdot \Delta_{out})$ time and $O(\Delta \cdot \min\{\Delta_{in}, \Delta_{out}\})$ space. The total time and space complexities for  computing all vertices' corenesses are $O(R_{SC} \cdot \Delta_{in}  \cdot m)$ and $O(\min\{\Delta_{in}, \Delta_{out}\} \cdot  m)$, respectively.}
\label{theorem:SCTC}
\end{theorem}

\begin{theorem}(\textbf{Message Complexity})
\gcc{The message complexity of Algorithm~\ref{algorithm:skylinecoreness} is $O(\Delta^2)$. The total message complexity for computing all vertices' corenesses is  $O(\Delta \cdot  m)$.}
\label{theorem:SCMC}
\end{theorem}

\vspace*{-.1cm}

%
%
%

Through the above analysis, we can see that the skyline coreness-based approach in Algorithm~\ref{algorithm:skylinecoreness} \emph{takes less space} and \emph{runs much faster} than the anchored coreness approach in Algorithm~\ref{algorithm:ACC}. 

\stitle{Block-centric extension.} Algorithm~\ref{algorithm:skylinecoreness} can be easily extended to the block-centric framework.
The only difference is that each machine iteratively computes the $n$-order D-index locally until the algorithm converges within the local block, before broadcasting to other blocks (lines 9-13 of Algorithm~\ref{algorithm:skylinecoreness}).


\section{Performance Evaluation}
\label{Sec:experiment}

In this section, we empirically evaluate 
our proposed algorithms. We conduct our
experiments on a collection of Amazon EC2 r5.2x large instances, each powered by 8 vCPUs and 64GB memory. 
The network bandwidth is up to 10G Gb/s. All experiments are implemented in C++ on the Ubuntu 18.04 operating system.


\vspace{0.1cm}
\noindent  \textbf{Datasets.} We use 11 real-world graphs
 in our experiments.
 Table~\ref{tabdataset} shows the statistics of these graphs.  Specifically, \texttt{Wiki-vote}\footnote{http://snap.stanford.edu/data/index.html\label{web}} is a voting graph;
\texttt{Email-EuAll}\textsuperscript{\ref{web}} is a communication graph;
\texttt{Amazon}\textsuperscript{\ref{web}} is a product co-purchasing graph;
\texttt{Hollywood}\textsuperscript{\ref{web2}} is an actors collaboration graph;
\texttt{Pokec}\textsuperscript{\ref{web}},
\texttt{Live Journal}\textsuperscript{\ref{web}}, and
\texttt{Slashdot}\textsuperscript{\ref{web}} are social graphs;
\texttt{Citation}\textsuperscript{\ref{web}} is a citation graph;
\texttt{UK-2002}\footnote{http://law.di.unimi.it/datasets.php\label{web2}},
\texttt{IT-2004}\textsuperscript{\ref{web2}}, and
\texttt{UK-2005}\textsuperscript{\ref{web2}} are web graphs.

\begin{table}[t]

\caption{Statistics of the datasets ($\bm{deg_{avg}}$ represents the average degree; K = $10^3$, M = $10^6$, and B = $10^9$)}
\vspace{-0.2cm}
\scalebox{0.9}{
\setlength\tabcolsep{3pt}
\begin{tabular}{|c|c|c|c|c|c|c|}
\hline
\textbf{Dataset} &\textbf{Abbr.} &$\bm{|V_{G}|}$ & $\bm{|E_{G}|}$ & $\bm{deg_{avg}}$ & $\bm{k_{max}}$ & $\bm{l_{max}}$  \\ \hline \hline
\texttt{Wiki-vote} & \texttt{WV} & 7.1\textbf{K} & 103.6\textbf{K} & 14.57 & 19 & 15  \\ \hline
\texttt{Email-EuAll} &\texttt{EE} & 265.2\textbf{K} & 420\textbf{K} & 1.58 & 28 & 28
\\ \hline
\texttt{Slashdot} &\texttt{SL} & 82.1\textbf{K} & 948.4\textbf{K} & 11.54 & 54 & 9  \\ \hline
\texttt{Amazon} & \texttt{AM} & 400.7\textbf{K} & 3.2\textbf{M} & 7.99 & 10 & 10  \\ \hline
\texttt{Citation} & \texttt{CT} & 3.7\textbf{M} & 16.5\textbf{M} & 4.37 & 1 & 1  \\ \hline
\texttt{Pokec} & \texttt{PO} &1.6\textbf{M} & 30.6\textbf{M} & 18.75 & 32 & 31  \\ \hline
\texttt{Live Journal} &\texttt{LJ} & 4.8\textbf{M} & 69.0\textbf{M} & 14.23 & 253 &254   \\ \hline
\texttt{Hollywood} & \texttt{HW} &2.1\textbf{M} & 228.9\textbf{M} & 105.00 & 1,297 & 99  \\ \hline
\texttt{UK-2002} & \texttt{UK2} &18.5\textbf{M} & 298.1\textbf{M} & 16.09 & 942 & 99 \\
\hline
\texttt{UK-2005} & \texttt{UK5} &39.4\textbf{M} & 936.3\textbf{M} & 23.73 & 584 & 99  \\ \hline
\texttt{IT-2004} & \texttt{IT} &41.2\textbf{M} & 1.1\textbf{B} & 27.87 & 3,198 & 990  \\ \hline
\end{tabular}
}
\label{tabdataset}
\end{table}

\vspace{0.1cm}
\noindent  \textbf{Algorithms.}
We compare five algorithms in our experiments.
\begin{itemize}
  \item \textbf{AC-V} and \textbf{AC-B}: The distributed anchored coreness-based D-core decomposition algorithms implemented in the vertex-centric and block-centric  frameworks, respectively.
  \item \textbf{SC-V} and \textbf{SC-B}: The distributed skyline coreness-based D-core decomposition algorithms implemented in the vertex-centric and block-centric frameworks, respectively.
  \item \textbf{Peeling}: The distributed version of the peeling algorithm for D-core decomposition~\cite{dcore-cs}, in which one machine is assigned as the coordinator to collect global graph information and dispatch decomposition tasks.
\end{itemize}

We employ GRAPE~\cite{fan2017grape} as the block-centric framework 
 and use the hash partitioner for graph partitioning by default. For the sake of fairness, we also employ GRAPE to simulate the vertex-centric framework. In specific, at each round, all vertices within a block execute computations only once and when all vertices complete the computation, the messages will be broadcast to their neighbors.

\vspace{0.1cm}
\stitle{Parameters and Metrics.} The parameters tested in  experiments include $\#$ machines and graph size, whose default settings are 8 and $100\% \cdot |V_G|$, respectively. The performance metrics evaluated include $\#$ iterations required for convergence, convergence rate (i.e., the percentage of vertices who have computed the coreness), running time (in seconds), and communication overhead (i.e., the total messages sent by all vertices).

\subsection{Convergence Evaluation }

The first set of experiments evaluates the convergence of our proposed algorithms.

\begin{table}[t]
\caption{$\#$ Iterations required for the algorithms}
\label{tab:noofite}
\vspace{-0.2cm}
\begin{tabular}{|c|c|c|c|c|c|c|}
\hline
\multicolumn{2}{|c|}{\multirow{2}{*}{\textbf{Algorithms}}} & \multicolumn{5}{c|}{\textbf{Datasets}} \\ \cline{3-7}
\multicolumn{2}{|c|}{}                         & \texttt{WV}  & \texttt{EE}  & \texttt{SL}   & \texttt{AM}   & \texttt{CT}  \\ \hline
\multicolumn{2}{|c|}{Upper Bound}           &  1,167  &  7,636  &  5,064  &  2,757   & 793   \\ \hline
\multirow{4}{*}{AC-V}   & Phase I  & 19  & 17  & 40   & 16   & 32  \\ \cline{2-7}
                                     & Phase II  & 32  & 19  & 53   & 64   & 32  \\ \cline{2-7}
                                     & Phase III  & 33  & 22  & 61   & 61   & 2   \\ \cline{2-7}
                                     & Total     & 84  & 58  & 154  & 141  & 66  \\ \hline
\multirow{4}{*}{AC-B}   & Phase I  & 14   & 14  & 35   & 13  & 28  \\ \cline{2-7}
                                     & Phase II  & 15   & 7   & 43   & 30  & 28  \\ \cline{2-7}
                                     & Phase III  & 16   & 21  & 45   & 25  & 2   \\ \cline{2-7}
                                     & Total     & 45   & 42  & 123  & 68  & 58  \\ \hline
\multicolumn{2}{|c|}{SC-V}           & 33  & 19  & 61   & 65   & \textbf{2}   \\ \hline
\multicolumn{2}{|c|}{SC-B}           & \textbf{17}   &\textbf{6}   & \textbf{46}   & \textbf{25}  & \textbf{2}   \\ \hline
\end{tabular}
\end{table}

\begin{figure}[t]

\centering
\setlength{\abovecaptionskip}{-1pt}
\subfigcapskip=-4pt
\subfigure[Phase I of AC-V and AC-B]{
	\includegraphics[width = 0.22\textwidth]{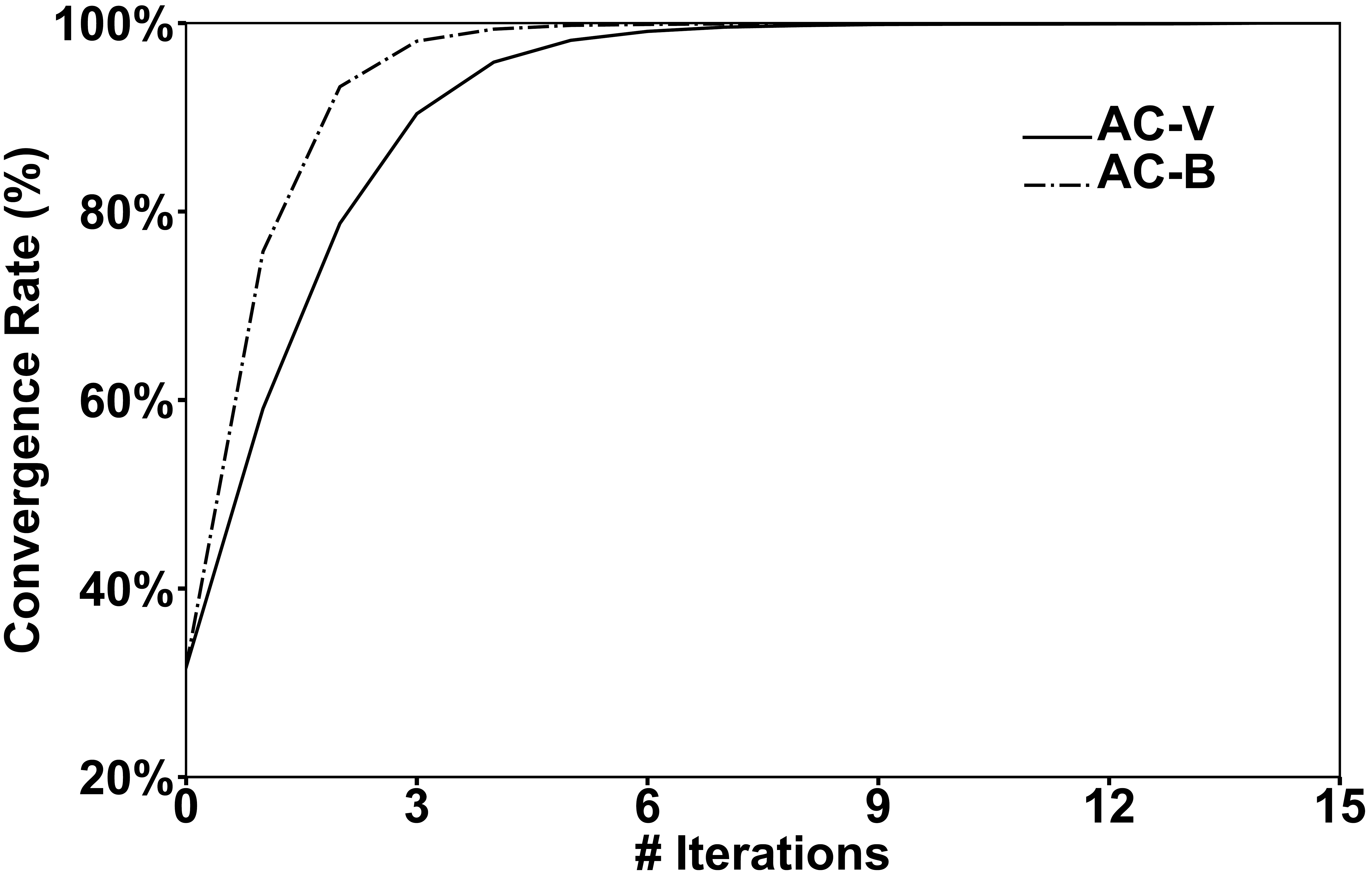}
}
\subfigure[Phase II of AC-V and AC-B]{
	\includegraphics[width = 0.22\textwidth]{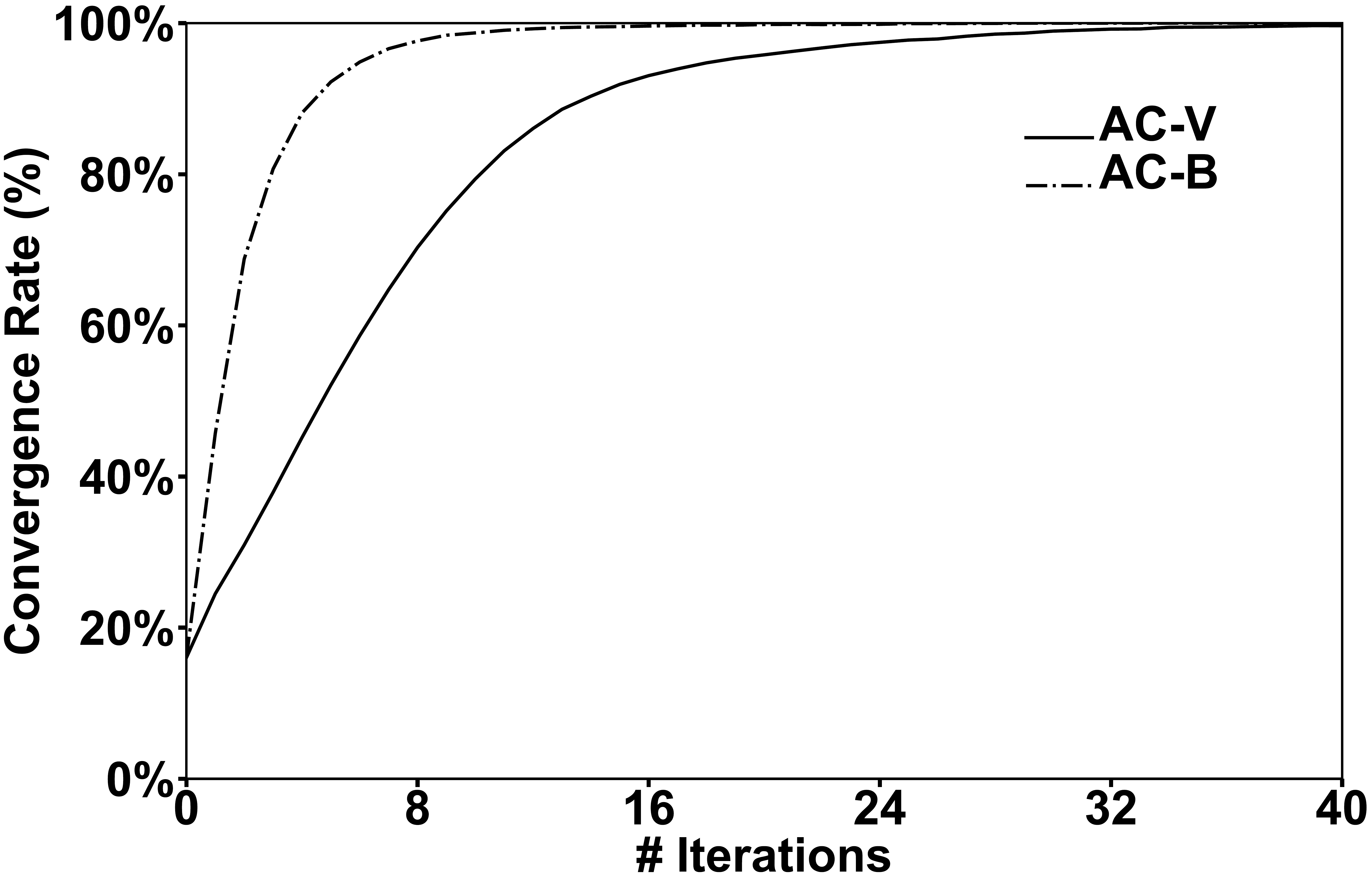}
}
\subfigure[Phase III of AC-V and AC-B]{
	\includegraphics[width = 0.22\textwidth]{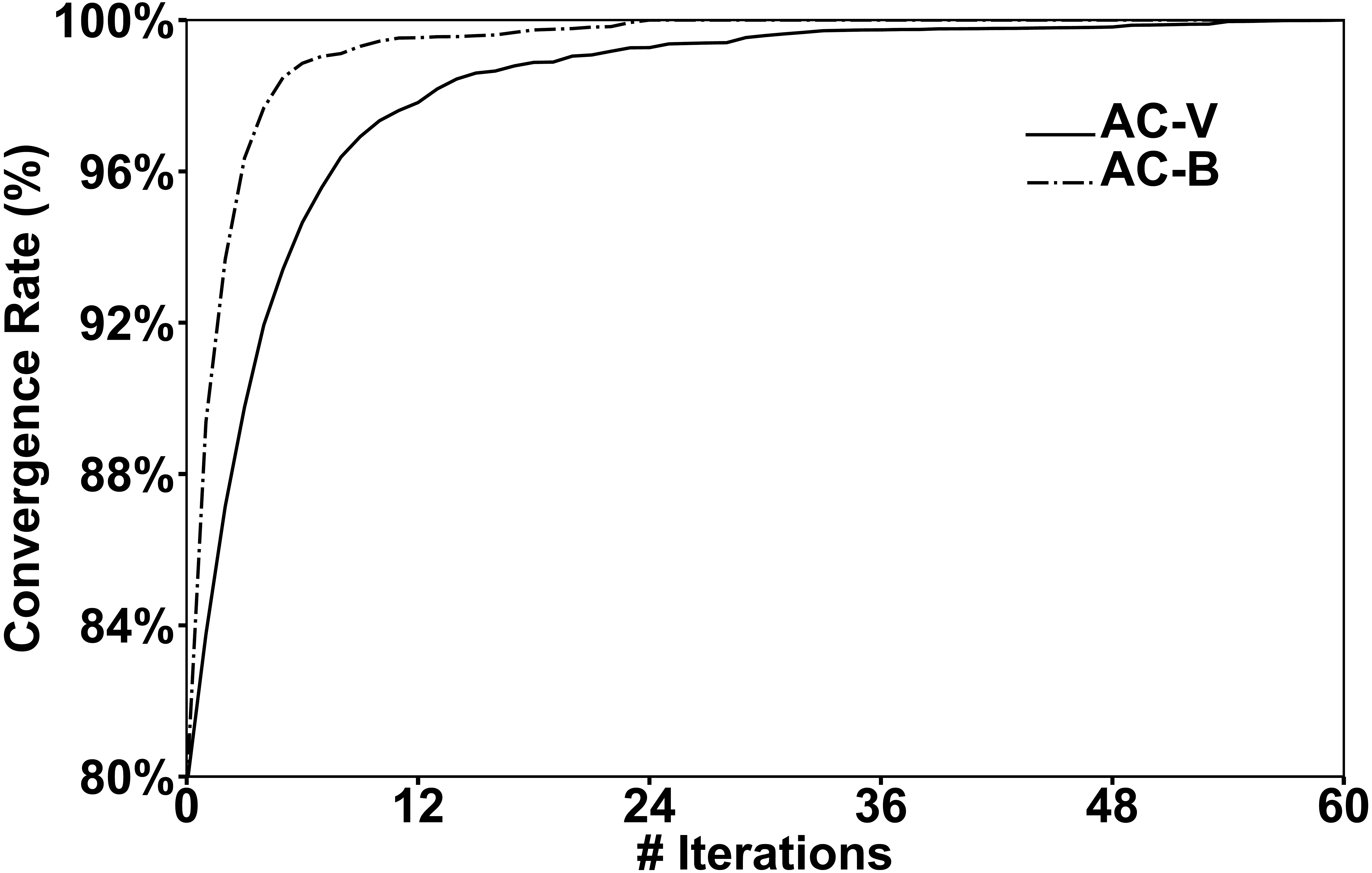}
}
\subfigure[SC-V and SC-B]{
\includegraphics[width = 0.22\textwidth]{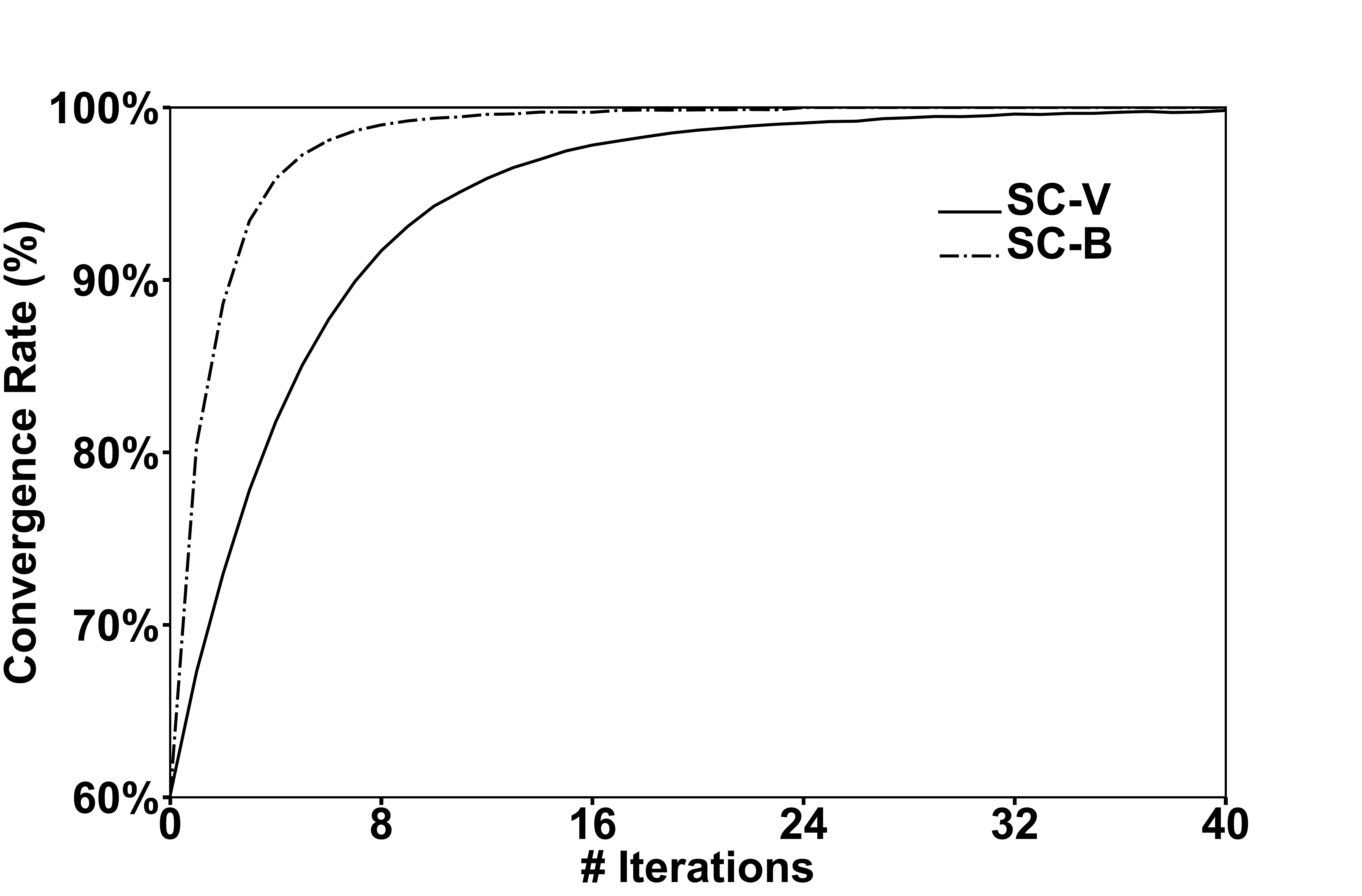}
\label{fig:exp2a}
}
\caption{Convergence rates of our algorithms (\texttt{AM})}
\label{fig:exp2}
\vspace{-0.6cm}
\end{figure}


\vspace{0.1cm}
\noindent \textbf{Exp-1: Evaluation on the number of iterations}.
We start by evaluating $\#$  iterations required for our algorithms to converge. Note that an iteration here refers to a cycle of the algorithm receiving messages, performing computations, and broadcasting messages. Table~\ref{tab:noofite} reports the results on datasets \texttt{WV}, \texttt{EE}, \texttt{SL}, \texttt{AM}, and \texttt{CT}. We make several observations.  First, for every graph, all of our proposed algorithms have much less iterations than the upper bound (i.e., the maximum degree of the graph), which demonstrates the efficiency of our algorithms. Second, the iterations of SC-V and SC-B are less than those of AC-V and AC-B. This is because the computation of anchored corenesses  is more cumbersome than that of skyline corenesses.  Hence, both AC-V and AC-B take more iterations. Third, for the same type of algorithms, i.e., AC or SC,  the algorithm implemented in the block-centric framework takes less iterations than that in the vertex-centric framework. The reason is that the block-centric framework allows algorithms to use vertices located in the same block to converge locally  within a single round, which leads to faster convergence.

\vspace{0.1cm}
\noindent \textbf{Exp-2: Evaluation on the convergence rate}.
\gcc{Since different vertices require different numbers of iterations to converge, in this experiment, we 
evaluate the algorithms' convergence rates. Figure~\ref{fig:exp2} shows the results on \texttt{Amazon}.  As expected, the algorithms implemented in the block-centric framework converge faster. For example, in Figure~\ref{fig:exp2a}, after 8 iterations, the convergence rates of SC-V and SC-B reach $89.9\%$ and $98.6\%$, respectively. Moreover, most vertices can converge within just a few iterations. 
Specifically, for SC-B, more than $95\%$ vertices converge within 5 iterations.
 In addition, SC algorithms have faster convergence rates than AC algorithms.  For example, AC-B takes 68 
iterations  to reach convergence while SC-B takes 25 iterations.}


\subsection{Efficiency Evaluation}
\label{exp:efficiency}
Next, we evaluate the efficiency of our proposed algorithms against the state-of-the-art peeling algorithm, denoted as Peeling. Note that if an algorithm cannot finish within 5 days, we denote it by 'INF'.

\begin{figure}[t]
\vspace{-0.2cm}
\centering
\setlength{\abovecaptionskip}{-4pt}
\subfigcapskip=-6pt
\subfigure[Running time]{\label{fig:exp4t}
\includegraphics[width = 0.4\textwidth]{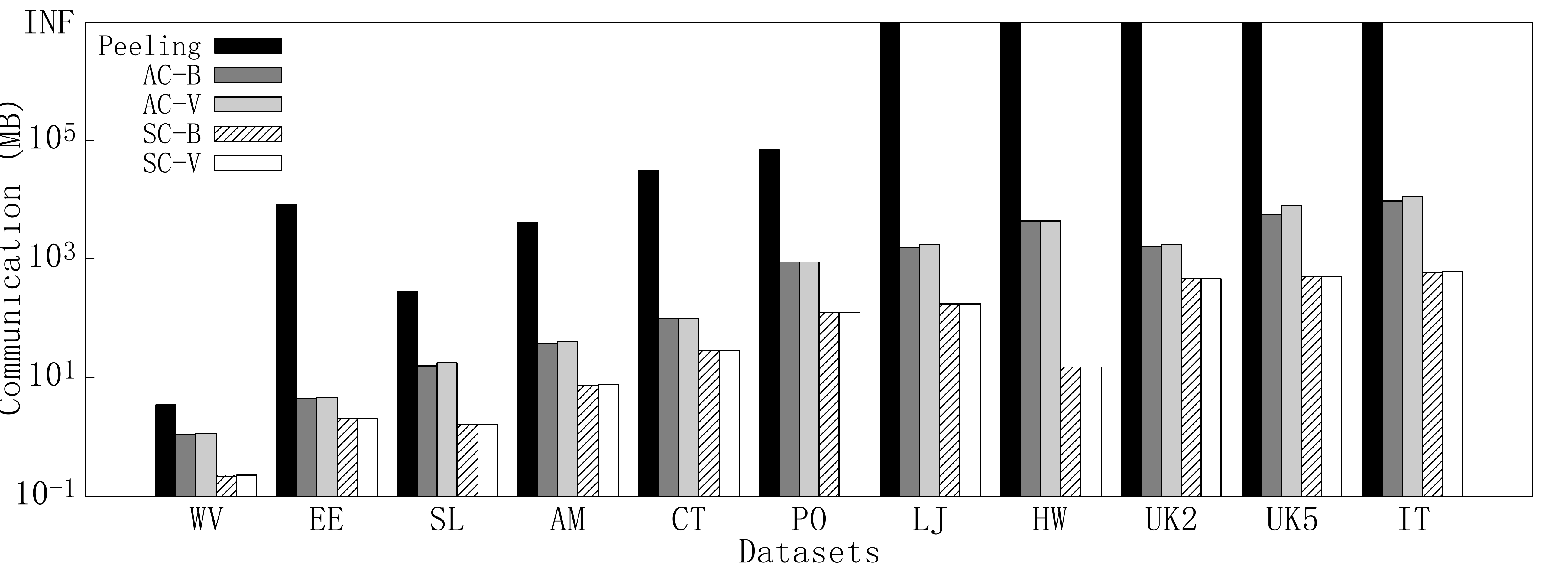}
}\vspace{-0.4cm}
\subfigure[Communication]{\label{fig:exp4m}
\includegraphics[width = 0.4\textwidth]{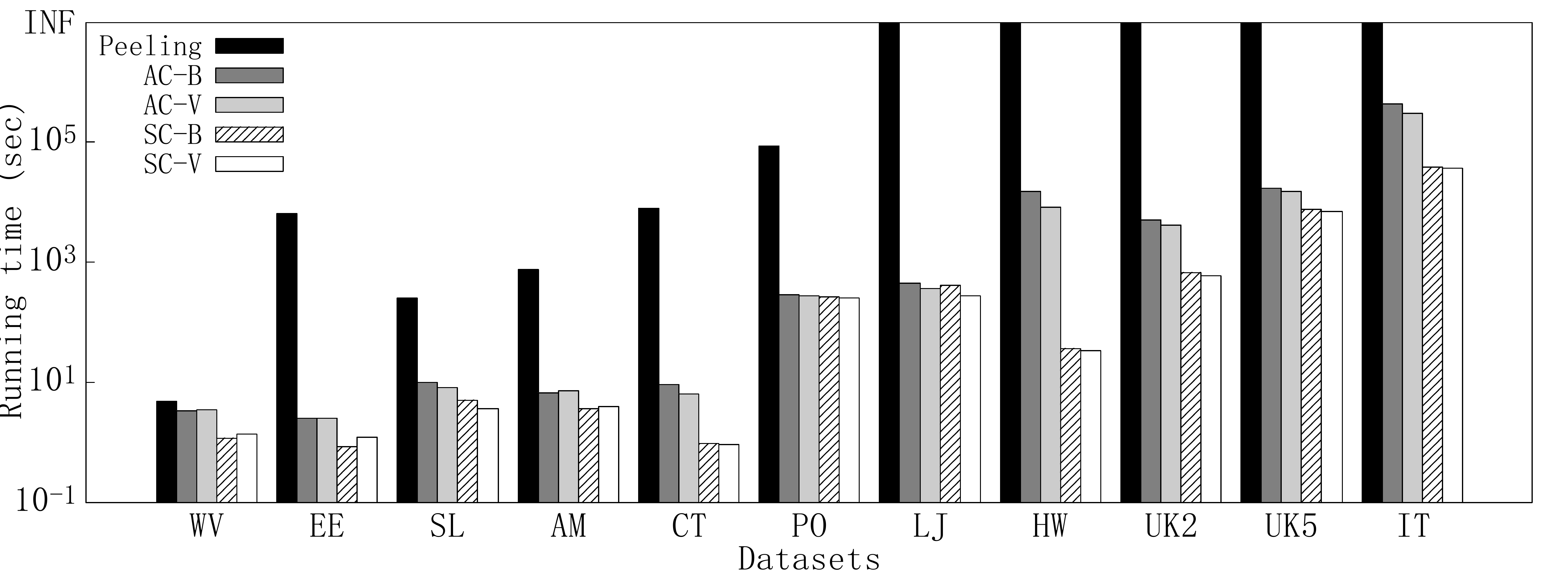}
}
\caption{Performance comparisons}
\label{fig:exp4}
\vspace{-0.3cm}
\end{figure}

\vspace{0.1cm}
\noindent \textbf{Exp-3: Our algorithms vs. Peeling.}
We first compare the  performance of our proposed algorithms with Peeling under the default experiment settings.
Figures~\ref{fig:exp4t} and~\ref{fig:exp4m} report the results in terms of the running time and communication overhead, respectively. First, we can see that Peeling cannot finish within 5 days on the large-scale graphs with more than 50 million edges, including \texttt{LJ}, \texttt{HW}, \texttt{UK2} \texttt{UK5}, and \texttt{IT}, while our algorithms can finish within 1 hour for most of these datasets and no more than 10 hours on the largest billion-scale graph for our fastest algorithm. Moreover, for the datasets where Peeling can finish, our algorithms outperform Peeling by up to 3 orders of magnitude. This well demonstrates the efficiency of our proposed algorithms. Second, SC-V and SC-B perform better than AC-V and AC-B in terms of both the running time and communication overhead. \gc{For example, on the biggest graph \texttt{IT} with over a billion edges, the improvement is nearly 1 order of magnitude.} This is because SC-V and SC-B compute less corenesses than AC-V and AC-B. Third, AC-V (resp. SC-V) is better than AV-B (resp. SC-B) in terms of the  running time while it is opposite in terms of the communication overhead. This is due to the effect of \emph{straggler}~\cite{osdiAnanthanarayananKGSLSH10}. Specifically, for the block-centric framework, in each iteration, the algorithms use block information to converge locally (i.e., within each block); they cannot start the next iteration until all blocks have converged. There are machines where 
some blocks may converge very slowly, which deteriorates the overall performance of the block-centric algorithms.


%

\begin{figure}[t]
\vspace{-0.2cm}
\centering
\setlength{\abovecaptionskip}{-4pt}
\subfigcapskip=-4pt
\subfigure[Running time (\texttt{HW})]{\label{fig:exp6a}
\includegraphics[width = 0.22\textwidth]{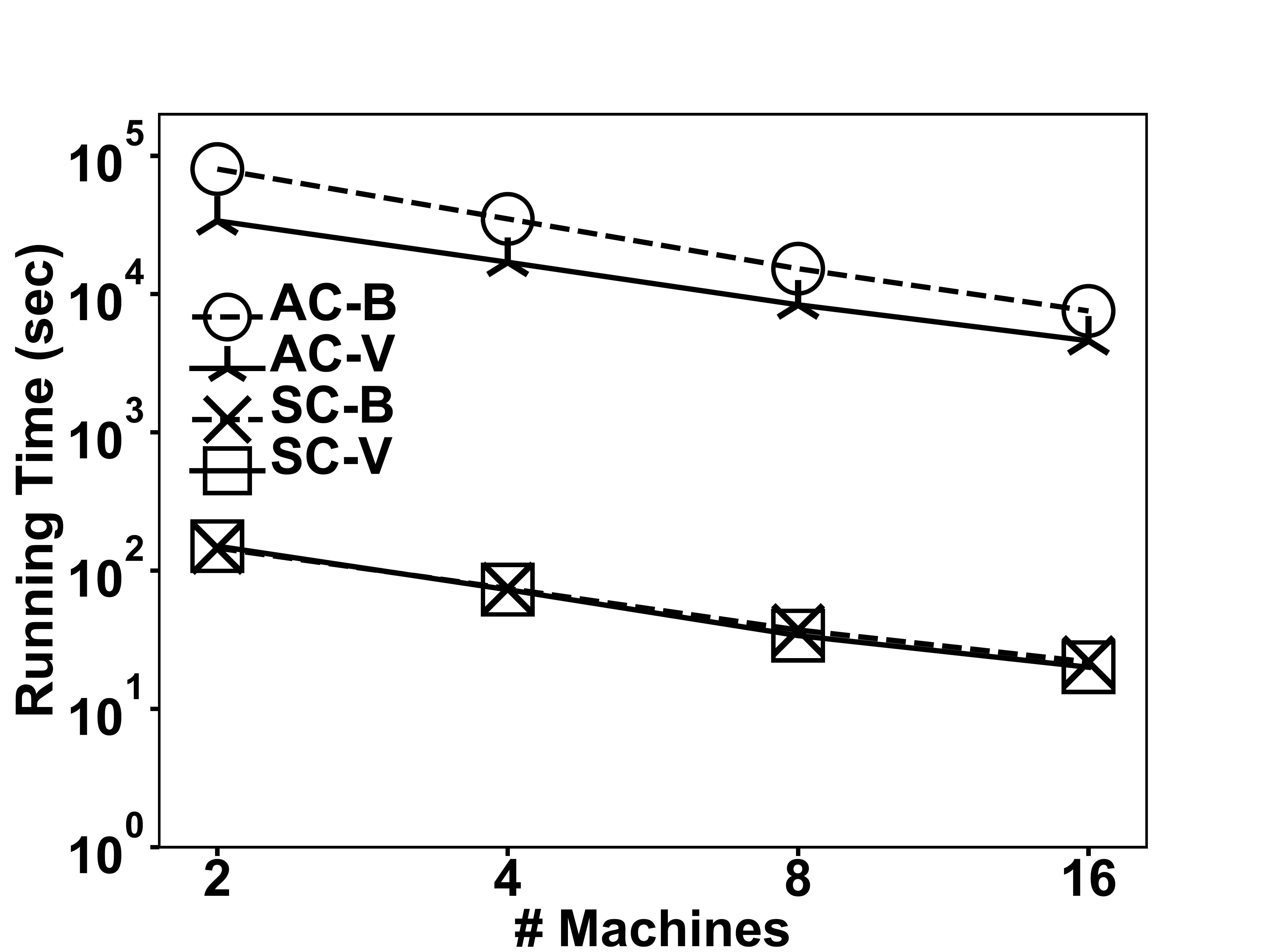}
}
\subfigure[Running time (\texttt{UK2})]{\label{fig:exp6b}
\includegraphics[width = 0.22\textwidth]{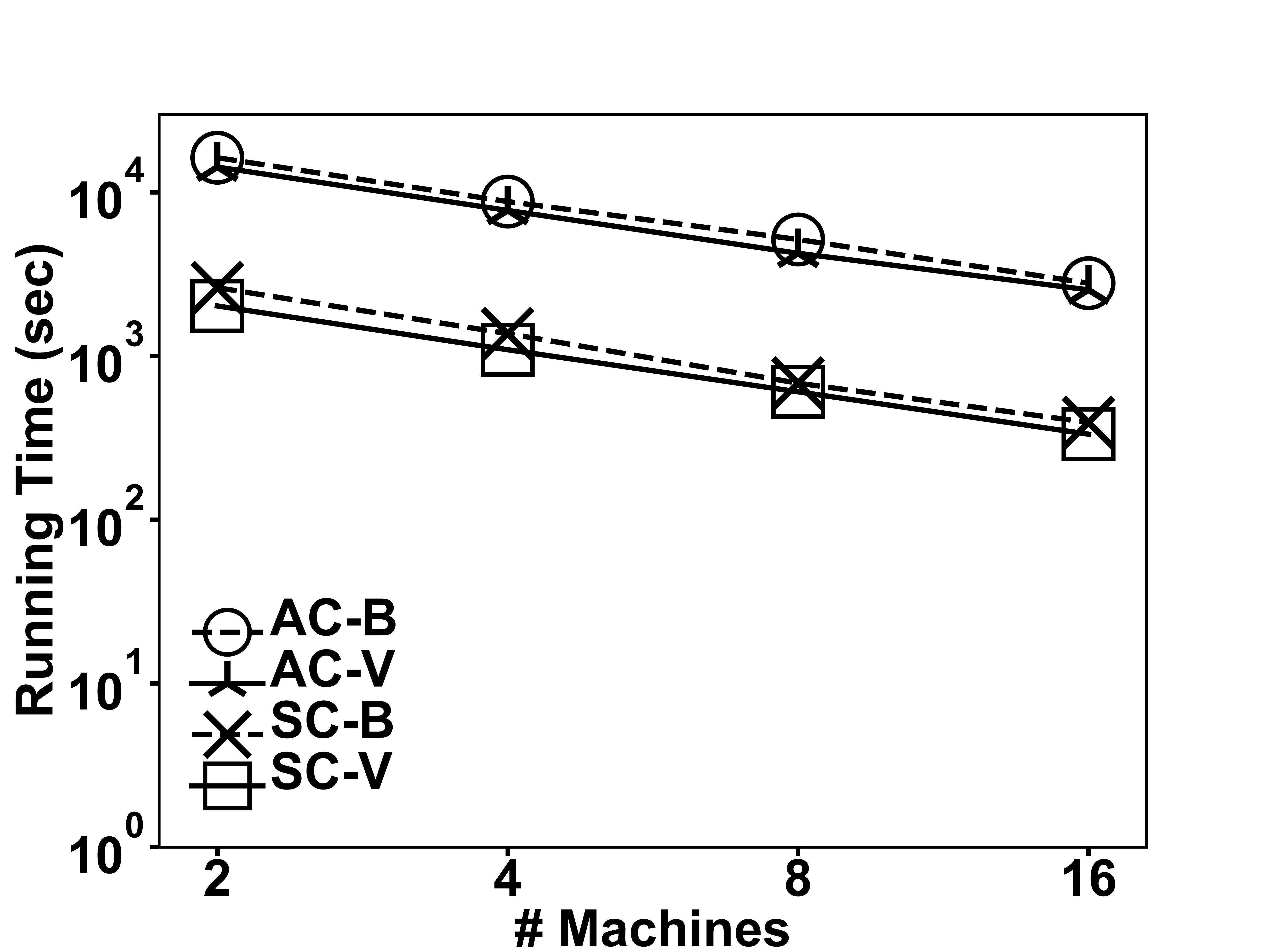}
}\vspace{-0.4cm}
\subfigure[Communication (\texttt{HW})]{\label{fig:exp6c}
\includegraphics[width = 0.22\textwidth]{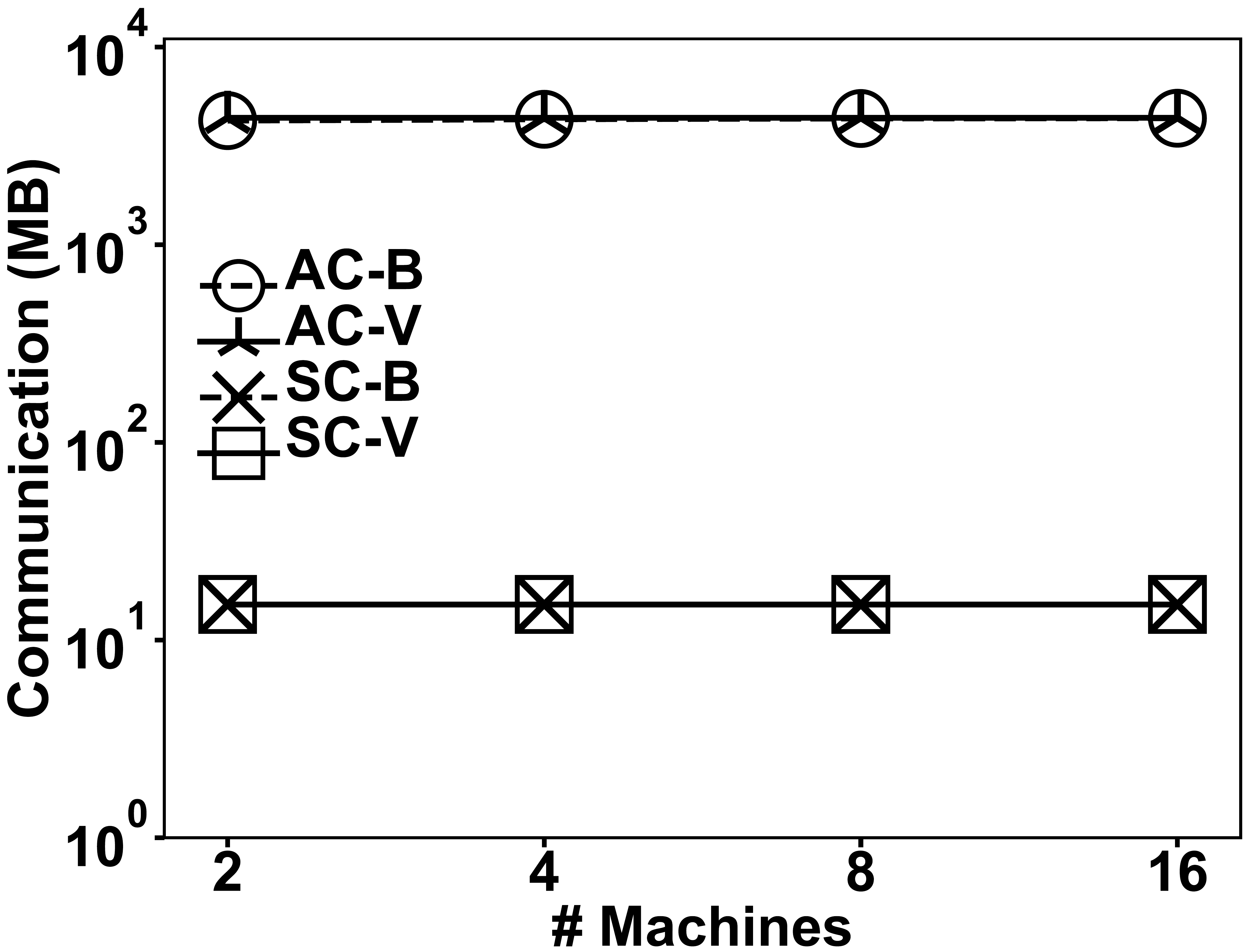}
}
\subfigure[Communication (\texttt{UK2})]{\label{fig:exp6d}
\includegraphics[width = 0.22\textwidth]{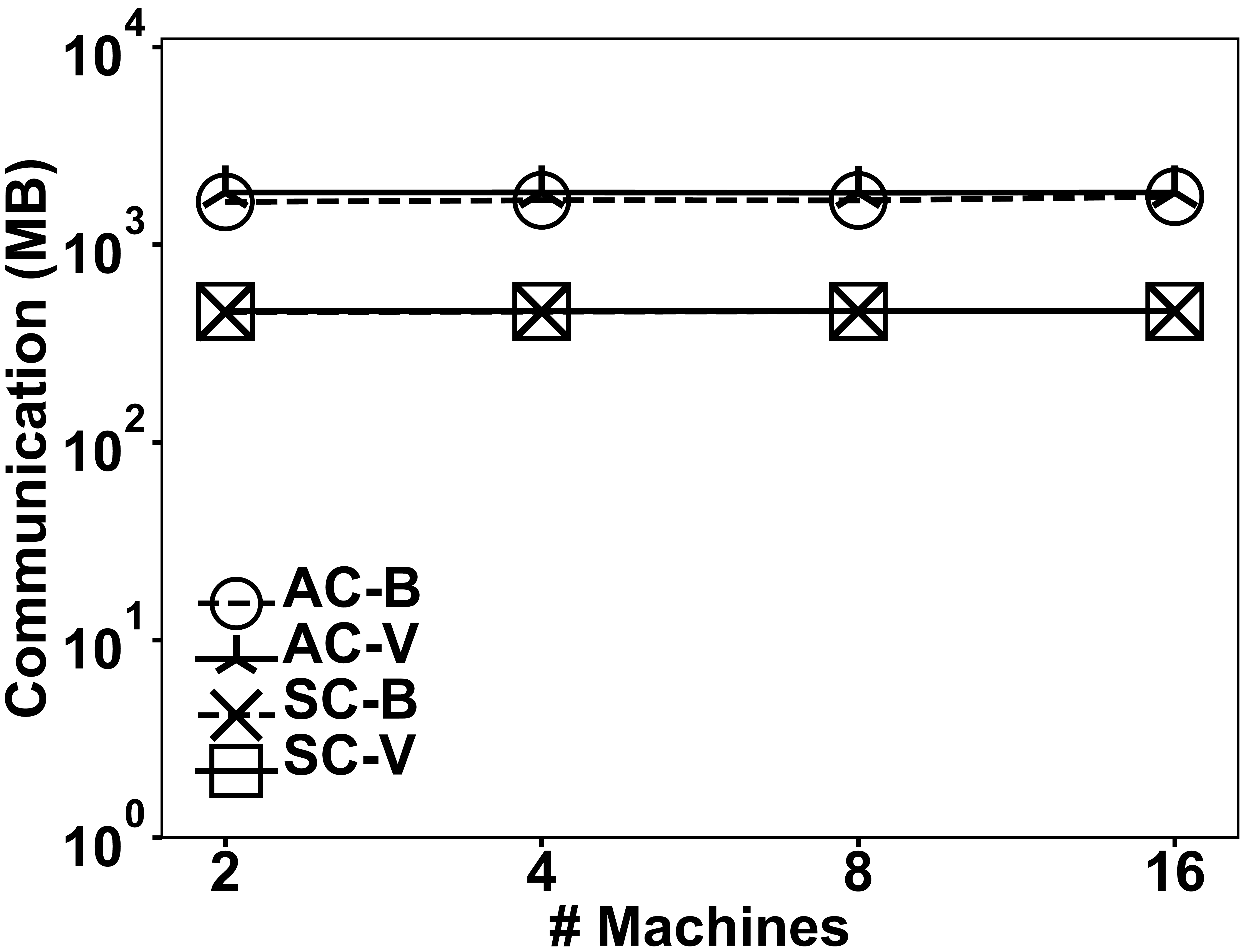}
}
\caption{Effect of $\#$ machines}
\label{fig:exp6}
\vspace{-0.4cm}
\end{figure}

\vspace{0.1cm}
\noindent \textbf{Exp-4: Effect of the number of  machines.}
Next, we vary the number of machines from 2 to 16 and test its effect on performance. Figure~\ref{fig:exp6} reports the results for datasets \texttt{UK2} and \texttt{HW}. As shown in Figures~\ref{fig:exp6a} and~\ref{fig:exp6b}, all of our algorithms take less running time when the number of machines increases.
This is because the more the machines, the stronger the computing power our algorithms can take advantage of, thanks to their distributed designs. In addition, Figures~\ref{fig:exp6c} and~\ref{fig:exp6d} show that the communication overheads of all algorithms do not change with the number of machines. This is because the communication overhead is determined by the convergence rate of the algorithms, which is not influenced by the number of machines.

\begin{figure}[t]
\centering
\setlength{\abovecaptionskip}{-4pt}
\subfigcapskip=-4pt
\subfigure[Running time (\texttt{PO})]{\label{fig:exp5a}
\includegraphics[width = 0.22\textwidth]{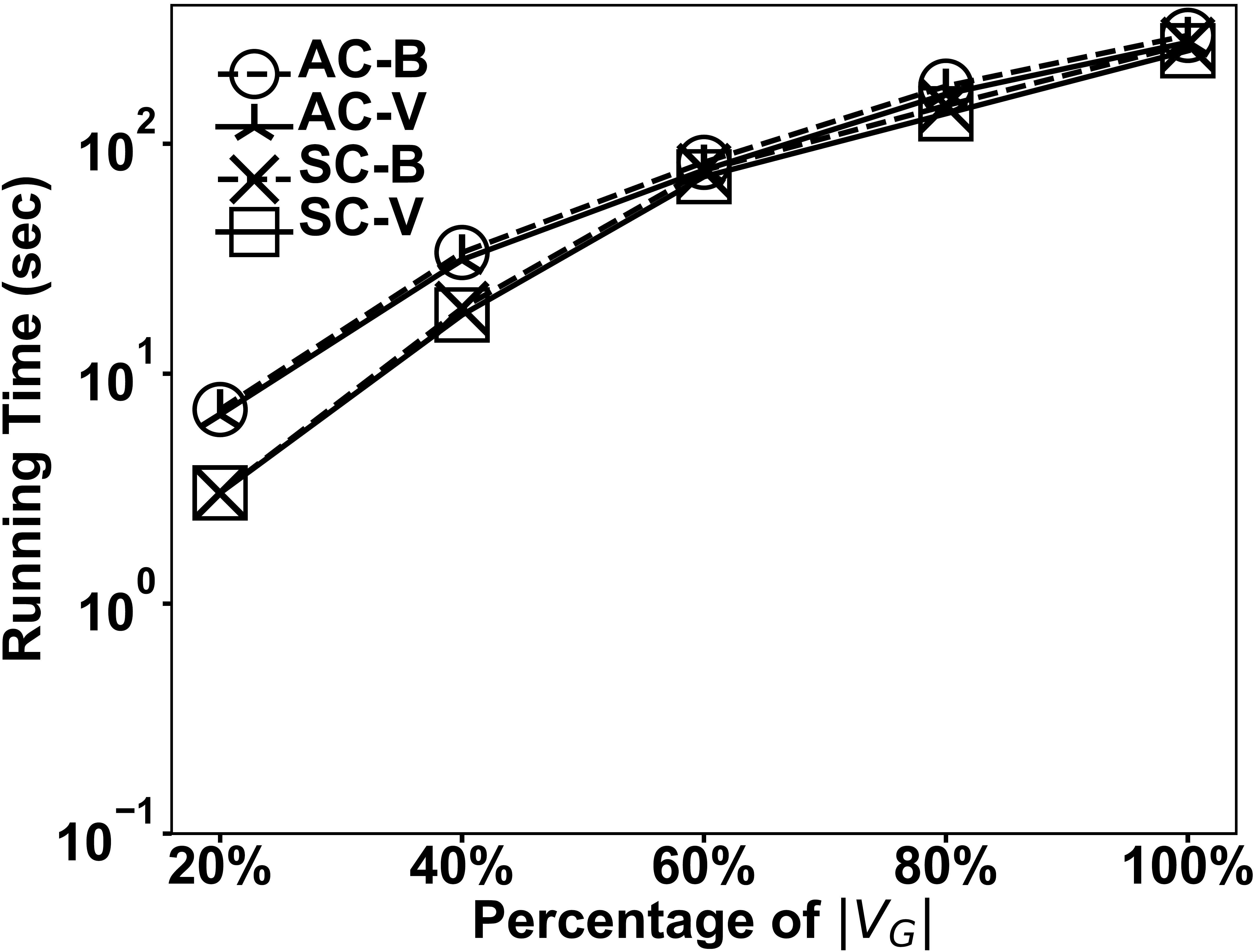}
}
\subfigure[Running time (\texttt{UK5})]{\label{fig:exp5b}
\includegraphics[width = 0.22\textwidth]{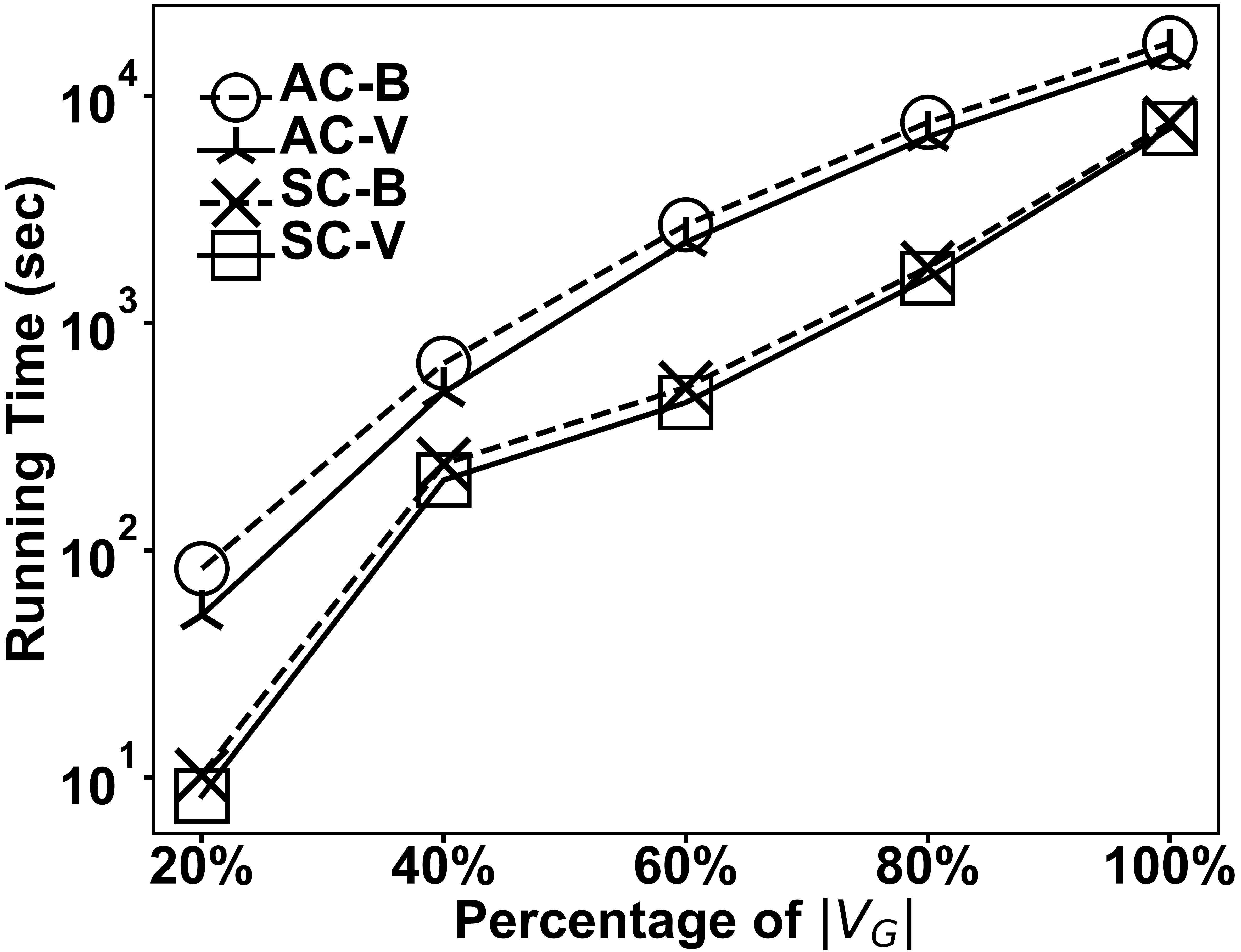}
}\vspace{-0.4cm}
\subfigure[Communication (\texttt{PO})]{\label{fig:exp5c}
\includegraphics[width = 0.22\textwidth]{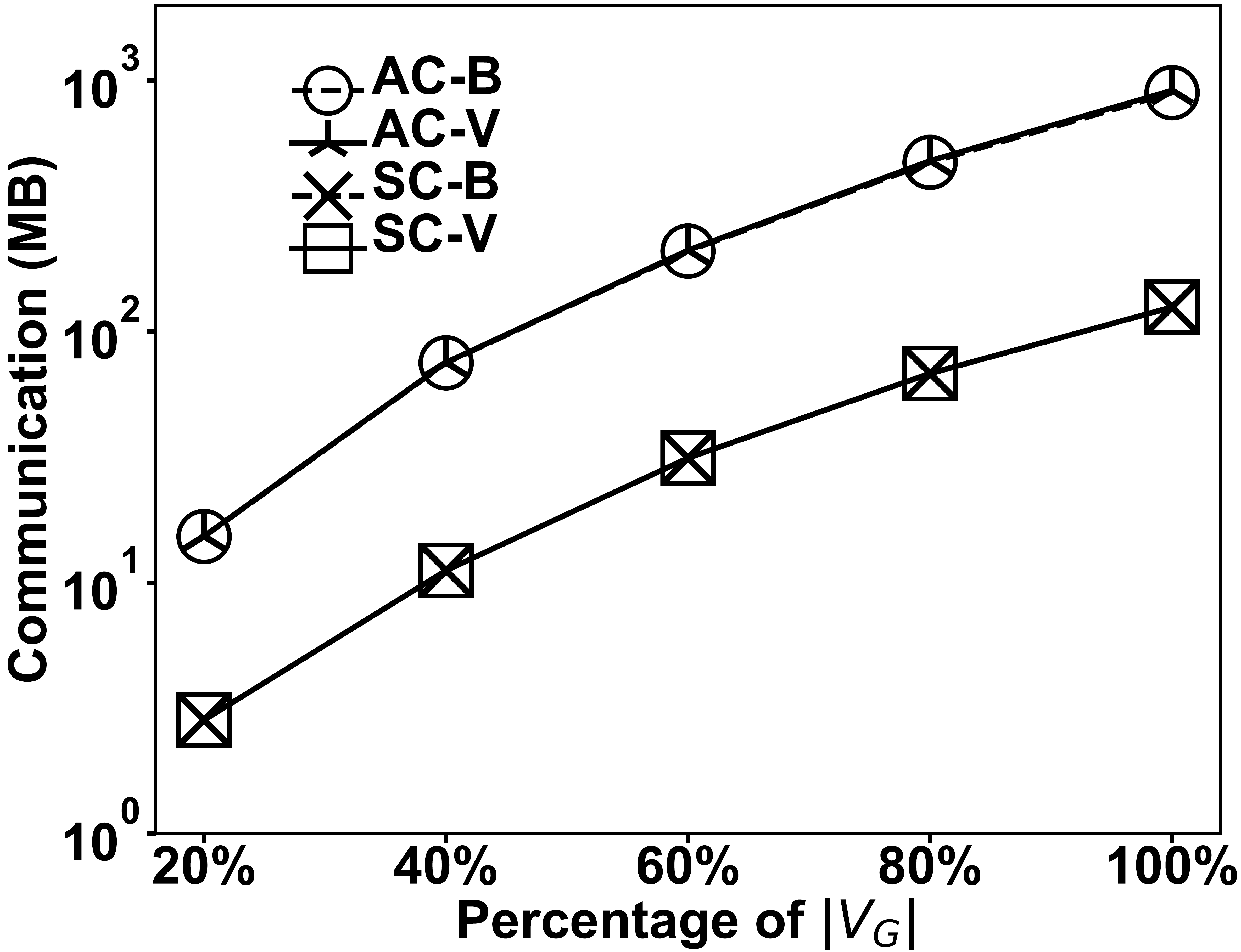}
}
\subfigure[Communication (\texttt{UK5})]{\label{fig:exp5d}
\includegraphics[width = 0.22\textwidth]{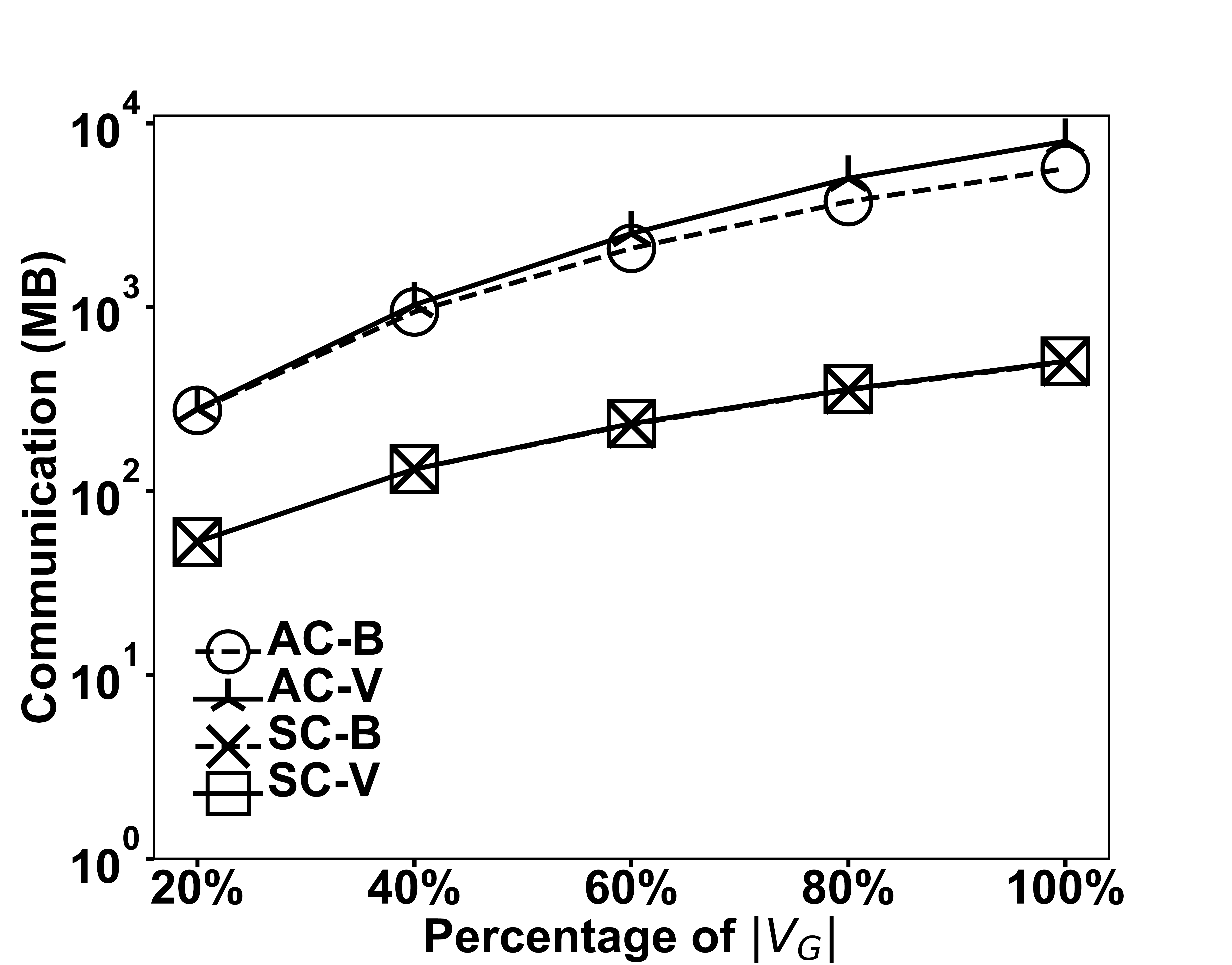}
}
\caption{Effect of cardinality}
\vspace{-0.6cm}
\label{fig:exp5}
\end{figure}

\vspace{0.1cm}
\noindent \textbf{Exp-5: Effect of dataset cardinality. }
We evaluate the effect of cardinality for our proposed algorithms on datasets \texttt{PO} and \texttt{UK5}. For this purpose, we extract a set of subgraphs from the original graphs by randomly selecting different fractions of vertices, which varies from $20\%$ to $100\%$. As shown in Figure~\ref{fig:exp5}, both the running time and communication overhead increase with the dataset cardinality. This is expected because the larger the dataset, the more the corenesses of the vertices to compute, resulting in to poorer performance.

\vspace{0.1cm}
\noindent \textbf{Exp-6: Effect of partition strategies. }
\gcc{We evaluate the effect of different partition strategies in block-centric algorithms, i.e., AC-B and  SC-B. Specifically, we compare four partition strategies, including {SEG}~\cite{fan2017grape}, {HASH}~\cite{fan2017grape}, {FENNEL}~\cite{FENNEL-partitioner}, and {METIS}~\cite{METIS-partitioner}.}

\begin{itemize}
  \item \gcc{SEG is a built-in partitioner of GRAPE. Let $C$ be the maximum cardinality of partitioned subgraphs. For a vertex $v$ with its ID $v_{id} \in [0, n-1]$, $v$ is allocated to the $i$-th subgraph, where $i= v_{id}/C$.}
  \item \gcc{HASH is also a built-in partitioner of GRAPE. Let $N$ be the number of  partitioned subgraphs. For a vertex $v$ with its ID $v_{id} \in [0, n-1]$, $v$ is allocated to the $i$-th subgraph, where $i=v_{id}\%N$.}
  \item \gcc{FENNEL subsumes two popular heuristics to partition the graph: the folklore heuristic that places a vertex to the subgraph with the fewest non-neighbors, and the degree-based heuristic that uses different heuristics to place a vertex based on its degree.}
  \item \gcc{METIS is a popular edge-cut partitioner  that partitions the graph into subgraphs with minimum crossing edges.}
\end{itemize}

\begin{figure}[t]
\centering
\setlength{\abovecaptionskip}{-4pt}
\subfigcapskip=-4pt
	\subfigure[Running time (AC-B)]{\label{fig:partitionACt}
	\includegraphics[width=0.23\textwidth]{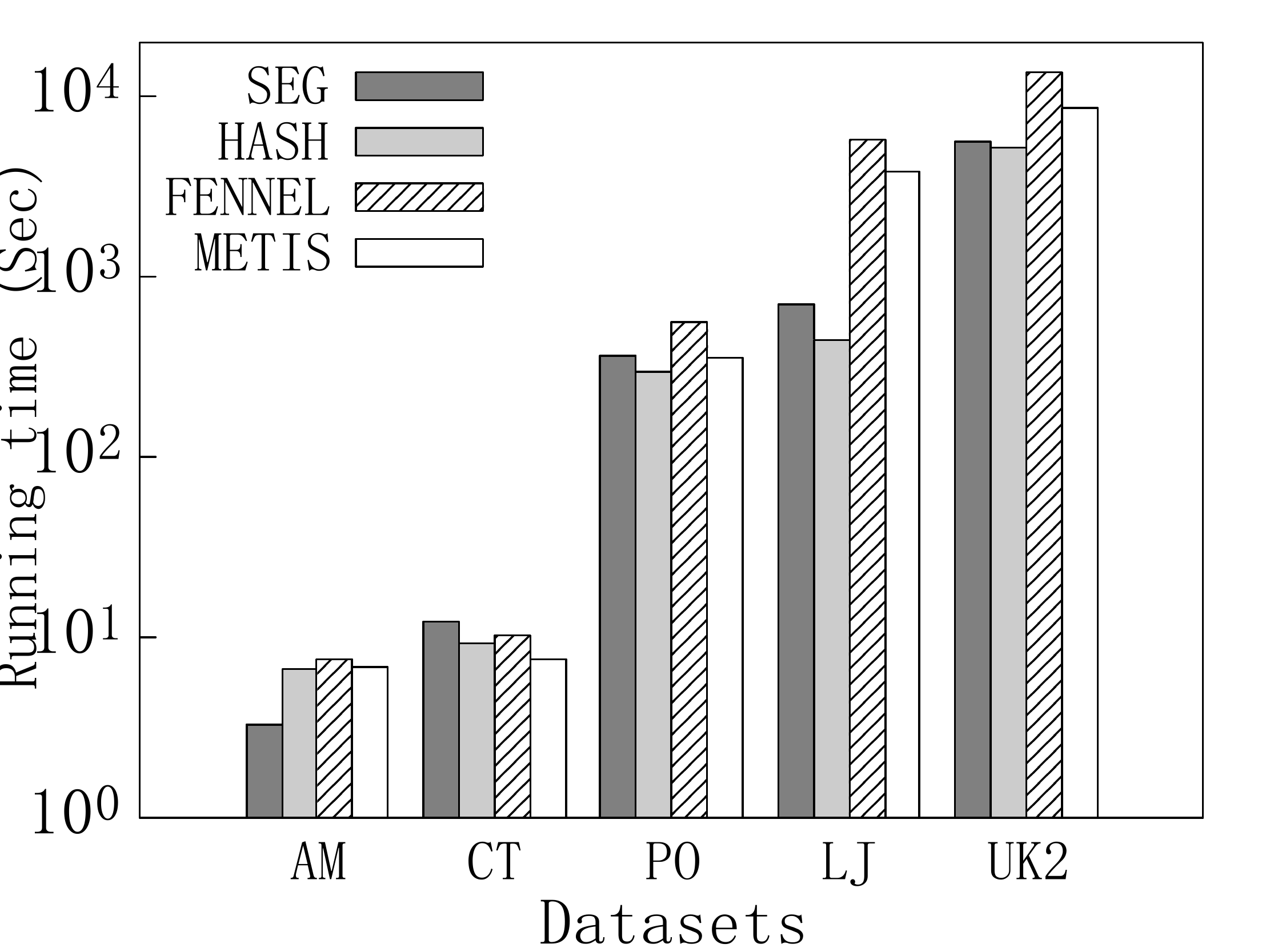}}
	\subfigure[Running time (SC-B)]{\label{fig:partitionSCt}
	\includegraphics[width=0.23\textwidth]{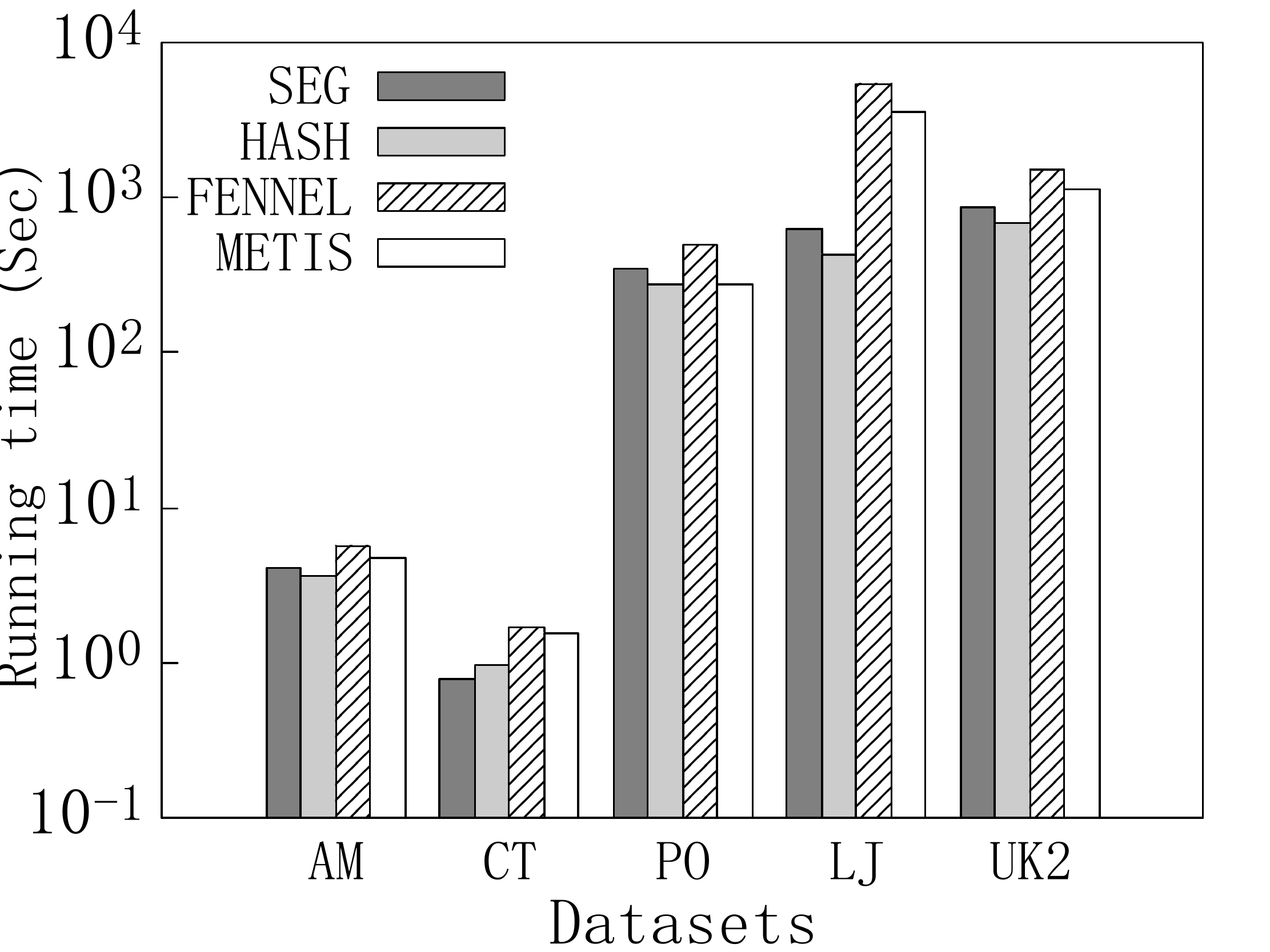}}\vspace{-0.4cm}
	\subfigure[Communication (AC-B)]{
		\includegraphics[width=0.23\textwidth]{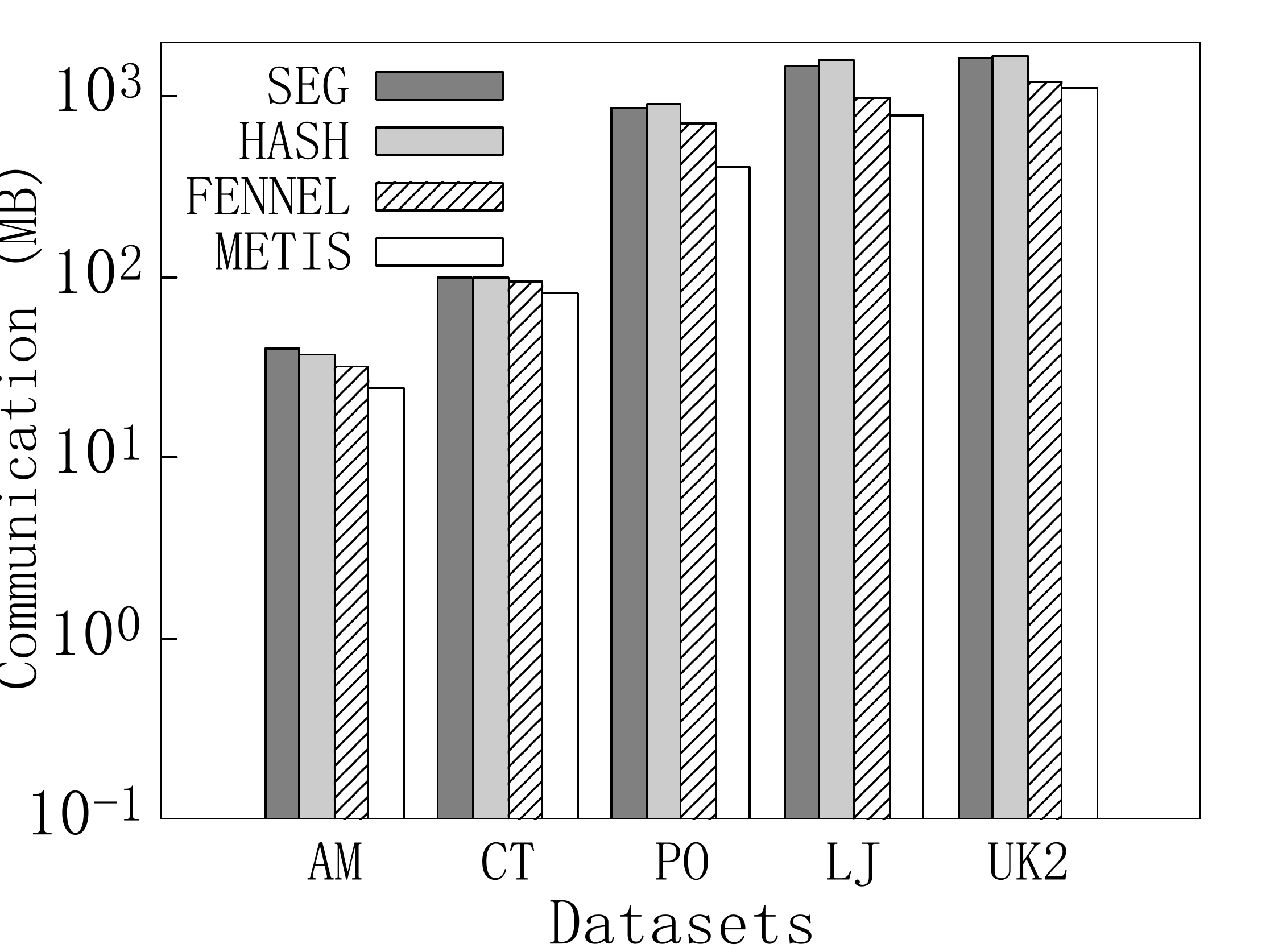}}
	\label{fig:partitionACc}
	\subfigure[Communication (SC-B)]{
		\includegraphics[width=0.23\textwidth]{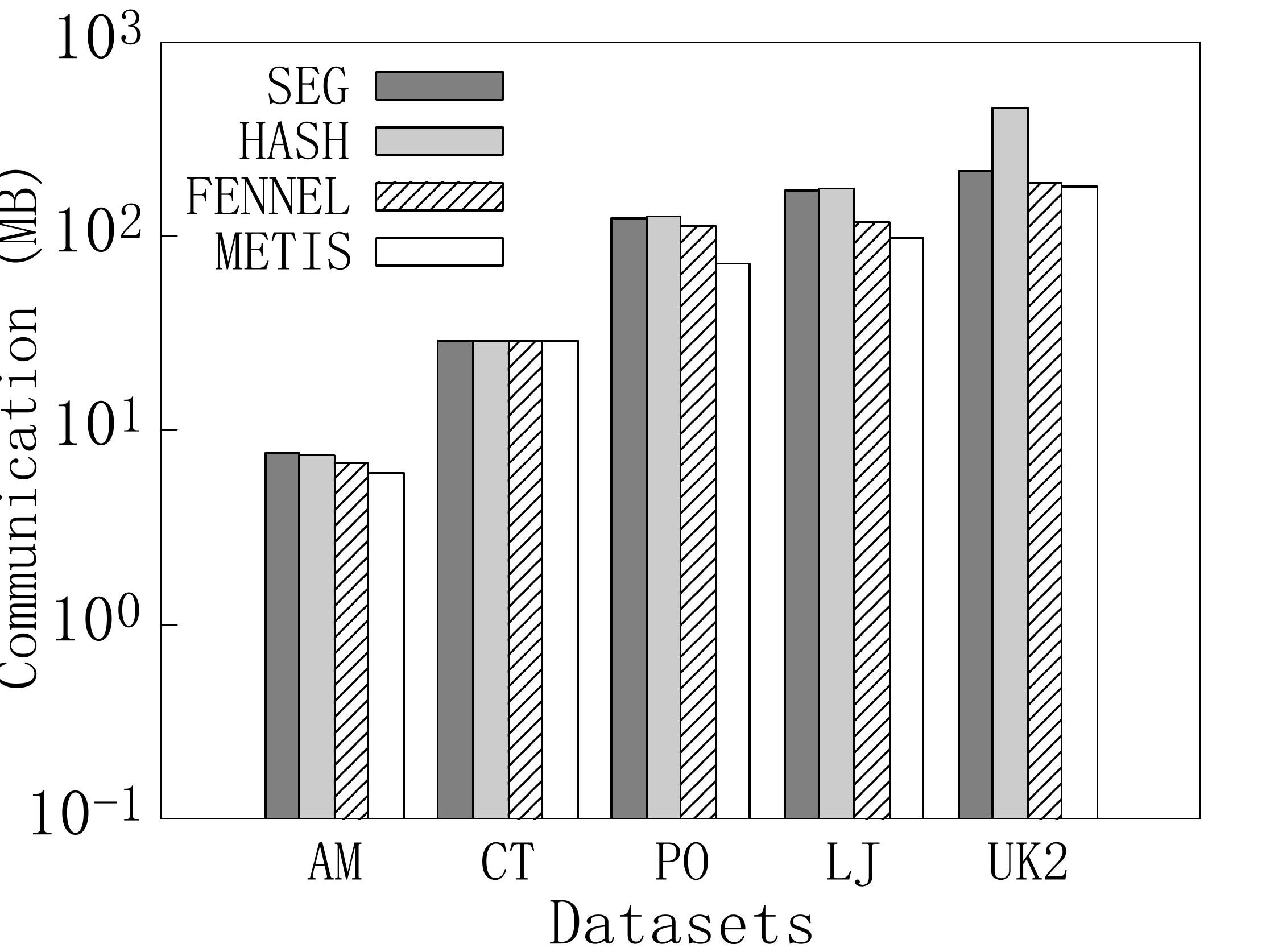}}
	\label{fig:partitionSCc}
	\caption{Effect of partition strategies}
	\label{fig:partition}
\vspace{-0.4cm}
\end{figure}

\gcc{Figure~\ref{fig:partition} shows the results. We can observe that HASH has the best performance in terms of running time on most datasets, but it has the worse performance in terms of communication cost. 
This is because HASH has more balanced partitions (i.e., each partition has almost  an equal number of vertices) while METIS and FENNEL have higher locality, which leads to more running time, due to the effect of \emph{straggler}, but less communication overhead. 
Considering the importance of efficiency in practice, we employ HASH as the default partition strategy in our experiments, as mentioned earlier.
}

\section{Conclusion }
\label{Sec:conclusion}

\gcc{In this paper, we study the problem of D-core decomposition in distributed settings. To handle distributed D-core decomposition, we propose two efficient algorithms, i.e., the anchored-coreness-based algorithm and skyline-coreness-based algorithm. Specifically, the anchored-coreness-based algorithm employs  in-H-index and out-H-index to compute the anchored corenesses in a distributed way; the skyline-coreness-based algorithm uses a newly designed index, called D-index, for D-core decomposition through skyline coreness computing. Both theoretical analysis and empirical evaluation show the efficiency of our proposed algorithms with fast convergence rates. }

\gcc{As for future work, we are interested to study how to further improve the performance of the skyline-coreness-based algorithm, in particular how to accelerate the  computation of D-index on each single machine. We are also interested to investigate efficient algorithms of distributed D-core maintenance for dynamic graphs.
}




\bibliographystyle{ACM-Reference-Format}
\bibliography{sample-base}

\appendix
\newpage

\section{Proof of Theorem~\ref{theorem:kmaxconvergence}}

\ \ \ \textsc{Theorem} 4.1. (Convergence).
\begin{equation*}
 k_{max}(v) = \lim_{n \rightarrow \infty} \mathrm{iH}_G^{(n)}(v)
\end{equation*}

\begin{proof}
First, we prove that  $\mathrm{iH}_G^{(n)}(v)$ is non-increasing with the increase of order $n$. Then, we prove that $k_{max}(v) \leq \mathrm{iH}_{G'}^{(\infty)}(v) \leq \mathrm{iH}_G^{(\infty)}(v)$, where $G' \subseteq G$ is a subgraph induced by the vertices $v'$ with $k_{max}(v')\geq k_{max}(v)$. Also, we know $k_{max}(v) \geq \mathrm{iH}_G^{(\infty )}(v)$ by definition. Hence, $k_{max}(v) = \mathrm{iH}_G^{(\infty)}(v)$. Complete proof is given here:
  	
First, we prove that  $\mathrm{iH}_G^{(n)}(v)$ is non-increasing with the increase of  $n$ through mathematical induction. (1) It is straightforward that $\mathrm{iH}_G^{(0)}(v)$ $\geq$ $\mathrm{iH}_G^{(1)}(v)$ holds according to the definition of $\mathrm{iH}_G^{(n)}(v)$. (2) Assume that $\mathrm{iH}_G^{(m)}(v) \geq \mathrm{iH}_G^{(m+1)}(v)$ holds. We have $\mathrm{iH}_G^{(m+2)}(v) =\mathcal{H}(\mathrm{iH}_G^{(m+1)}(u_{1}),$ $...,\mathrm{iH}_G^{(m+1)}(u_{k_{i}})) \leq$ $ \mathcal{H}(\mathrm{iH}_G^{(m)}(u_{1}), ... ,\mathrm{iH}_G^{(m)}(u_{k_{i}})) = \mathrm{iH}_G^{(m+1)}(v) $, i.e., $\mathrm{iH}_G^{(m+1)}(v) \geq \mathrm{iH}_G^{(m+2)}(v)$. Thus, if $\mathrm{iH}_G^{(n)}(v) \geq \mathrm{iH}_G^{(n+1)}(v)$ holds for $n=m$, it also holds for $n=m+1$.  Combining (1) and (2), $\mathrm{iH}_G^{(n)}(v)$ is non-increasing with the increase of  $n$. Moreover, since $\mathrm{iH}_G^{(n)}(v)$ is a positive integer, $\mathrm{iH}_G^{(n)}(v)$ can converge to a certain value.
  	
Next, we prove that $\mathrm{iH}_G^{(\infty)}(v) \geq k_{max}(v)$.  To this end, we introduce two  properties of $\mathrm{iH}_G^{(n)}(v)$: (i) if $G' \subseteq G$,   for $ \forall v \in V_{G'}$ and $\forall n \in \mathbb{N}_0$, we have $\mathrm{iH}_{G'}^{(n)}(v) \leq \mathrm{iH}_G^{(n)}(v)$; and (ii) for $\forall v \in V_G$ and $\forall n \in \mathbb{N}_0$, we have $\mathrm{iH}_G^{(n)}(v) \geq indeg_{min}(G)$, where $indeg_{min}(G)$ is the minimum in-degree of $G$. Let $G' \subseteq G$ be the subgraph of $G$ induced by vertex $v$ and the vertices $v'$ with $k_{max}(v')\geq k_{max}(v)$. We have $\mathrm{iH}_G^{(\infty)}(v) \geq \mathrm{iH}_{G'}^{(\infty)}(v) \geq indeg_{min}(G') \geq k_{max}(v)$.
  	
Then, we prove that $k_{max}(v) \geq \mathrm{iH}_G^{(\infty )}(v)$. To this end, we construct a subgraph $G'' \subseteq G$ induced by vertex $v$ and the vertices $v''$ satisfying $\mathrm{iH}_{G}^{(\infty)}(v'') \geq  \mathrm{iH}_{G}^{(\infty)}(v)$. Obviously, $G''$ is a $(\mathrm{iH}_{G}^{(\infty)}(v), 0)$-core. Hence, $k_{max}(v) \geq \mathrm{iH}_G^{(\infty )}(v)$.

Finally, combining the above three proofs, we  obtain Theorem 4.1.
\end{proof}

\section{Proof of Theorem~\ref{theorem:ACTC}}

\ \ \ \textsc{Theorem} 4.4. (\emph{\textbf{Time and Space Complexities}}).
\emph{Algorithm~\ref{algorithm:ACC} takes $O(R_{AC} \cdot \Delta_{in}  \cdot \Delta)$ time and $O(\Delta_{in}  \cdot \Delta)$ space. The total time and space complexities for  computing all vertices' corenesses are $O(R_{AC} \cdot \Delta_{in} \cdot m)$ and $O(\Delta_{in} \cdot m )$, respectively.}.

\begin{proof}
Algorithm~\ref{algorithm:ACC} consists of three phases in Algorithms~\ref{algorithm:kmax}, \ref{algorithm:lUPC}, and \ref{algorithm:refine}. Algorithm~\ref{algorithm:kmax} iteratively computes the $n$-order in-H-index for all vertices. The time complexity of  Algorithm~\ref{algorithm:kmax} is mainly determined by the number of rounds that need to compute the $n$-order in-H-index and also the processing time of each round, which are  $R_{AC-I}$ and $O(deg^{in}_G(v))$, respectively. Hence, the time complexity of Algorithm~\ref{algorithm:kmax} is $O(R_{AC-I} \cdot deg^{in}_G(v))$. Next, Algorithm~\ref{algorithm:lUPC} iteratively computes $l_{upp}(k, v)$. Each round of Algorithm~\ref{algorithm:lUPC} takes $O(deg^{in}_G(v) \cdot deg^{out}_G(v))$ time. Algorithm~\ref{algorithm:refine} iteratively refines $l_{upp}(k, v)$ to $l_{max}(k, v)$. Each round of Algorithm~\ref{algorithm:refine} takes $O(deg^{in}_G(v) \cdot deg_G(v))$ time.   Hence, the time complexities of Algorithms~\ref{algorithm:lUPC} and \ref{algorithm:refine} are $O(R_{AC-II} \cdot deg^{in}_G(v) \cdot deg^{out}_G(v))$ and $O(R_{AC-III} \cdot  deg^{in}_G(v) \cdot deg_G(v))$, respectively. Therefore, the time complexity of Algorithm~\ref{algorithm:ACC} is $O(R_{AC-I} \cdot deg^{in}_G(v) + R_{AC-II} \cdot deg^{in}_G(v) \cdot deg^{out}_G(v) + R_{AC-III} \cdot  deg^{in}_G(v) \cdot deg_G(v))$ $=$ $O(R_{AC} \cdot deg^{in}_G(v) \cdot deg_G(v))$ $=$ $O(R_{AC} \cdot \Delta_{in}  \cdot \Delta)$. The total time complexity for  all vertices is $O(\sum \limits_{v\in V_G} R_{AC} \cdot deg^{in}_G(v) \cdot deg_G(v))$ $=$ $O(R_{AC} \cdot \Delta_{in} \cdot \sum \limits_{v\in V_G} deg_G(v))$ $=$ $O(R_{AC} \cdot \Delta_{in} \cdot m)$.

Next, we analyze the space complexity. Algorithm~\ref{algorithm:kmax} computes $\mathrm{iH}_G^{(n)}(v)$ by storing $\mathrm{iH}_G^{(n-1)}(u)$ for $u \in N^{in}(v)$, which costs $O(deg^{in}_G(v))$ space. Next, Algorithm~\ref{algorithm:lUPC} computes $l_{upp}(k, v)$.  For each $k \in [0, k_{max}]$, it stores the out-H-indexes of the out-neighbors in $O(deg^{out}_G(v))$ space. Hence,  Phase II requires $O(deg^{in}_G(v) \cdot deg^{out}_G(v))$ space. Phase III refines $l_{upp}(k, v)$ to $l_{max}(k, v)$ in Algorithm~\ref{algorithm:refine}.  For each $k \in [0, k_{max}]$, it  stores $l_{upp}(k, v')$ for all $v$'s neighbors in $O(deg^{in}_G(v)\cdot deg_G(v))$ space. Overall, the space complexity of Algorithm~\ref{algorithm:ACC} is $O(deg^{in}_G(v) + deg^{in}_G(v) \cdot deg^{out}_G(v) + deg^{in}_G(v)\cdot deg_G(v))$ $=$ $O(deg^{in}_G(v)\cdot deg_G(v))$ $=$ $O(\Delta_{in} \cdot \Delta )$. The total space complexity for  all vertices is $O(\sum \limits_{v\in V_G} deg^{in}_G(v)\cdot deg_G(v))$ $=$ $O(\Delta_{in} \cdot  m)$.
\end{proof}

\section{Proof of Theorem~\ref{theorem:ACMC}}

\ \ \ \textsc{Theorem} 4.5. (\emph{\textbf{Message Complexity}}).
\emph{The message complexity (i.e., the total number of times that a vertex send messages) of Algorithm~\ref{algorithm:ACC} is $O(\Delta_{in}  \cdot \Delta_{out} \cdot \Delta)$. The total message complexity for computing all vertices' corenesses is  $O(\Delta_{in}  \cdot \Delta_{out} \cdot  m)$.}

\begin{proof}
We analyze the message complexity in terms of two parts: Algorithm~\ref{algorithm:kmax} and Algorithms~\ref{algorithm:lUPC} and~\ref{algorithm:refine}. First, Algorithm~\ref{algorithm:kmax} calculate $k_{max}(v)$ by iteratively computing $n$-order in-H-index of $v$. When $n$-order in-H-index is not equal to $(n-1)$-order in-H-index, Algorithm~\ref{algorithm:kmax} also sends messages to all $v$'s out-neighbors. $v$ sends message at most $deg^{in}_G(v)$ times. Hence, the message complexity of Algorithm~\ref{algorithm:ACC} is $O(deg^{in}_G(v) \cdot deg^{out}_G(v))$.


Next, for each value of $k$, Algorithm~\ref{algorithm:lUPC} first computes the upper bound  $l_{upp}(k, v)$ and then Algorithm~\ref{algorithm:refine}  refine $l_{upp}(k, v)$ to $l_{max}(k, v)$.  Algorithms~\ref{algorithm:lUPC} obtains $l_{upp}(k, v)$  by iteratively computing $n$-order out-H-index of $v$. When $n$-order out-H-index of $v$ decreases, Algorithm~\ref{algorithm:lUPC} sends messages to  $v$'s in-neighbors.  Algorithm~\ref{algorithm:lUPC} continues until the $n$-order out-H-index of $v$ equals $l_{upp}(k, v)$. Then, Algorithms~\ref{algorithm:refine} refines $l_{upp}(k, v)$ by gradually decreasing it to $l_{max}(k, v)$. Each time  $l_{upp}(k, v)$ decreases, Algorithm~\ref{algorithm:refine} sends messages to $v$'s neighbors. Since $l_{max}(k, v)$ is at least 1 and there are at most $deg^{in}_G(v)$ values of $k$, the total messages sent by $v$ is $deg^{out}_G(v)\cdot deg^{in}_G(v) \cdot deg_G(v) $ in  worst. Overall, the message complexity of Algorithms~\ref{algorithm:lUPC} and~\ref{algorithm:refine} is $O(deg^{out}_G(v) \cdot deg^{in}_G(v) \cdot deg_G(v))$.

As a result, the message complexity of  Algorithm~\ref{algorithm:ACC} is $O(deg^{in}_G(v) \cdot deg^{out}_G(v) + deg^{out}_G(v)  \cdot deg^{in}_G(v) \cdot deg_G(v))$ $=$ $O(deg^{out}_G(v)  \cdot deg^{in}_G(v) \cdot deg_G(v))$ $=$ $O(\Delta_{in}  \cdot \Delta_{out} \cdot \Delta)$. The total message complexity for all vertices is $O(\sum \limits_{v\in V_G} deg^{out}_G(v)  \cdot deg^{in}_G(v) \cdot deg_G(v))$ $=$ $O(\Delta_{in}  \cdot \Delta_{out} \cdot  \sum \limits_{v\in V_G}  deg_G(v))$ $=$ $O(\Delta_{in}  \cdot \Delta_{out} \cdot  m)$.
\end{proof}

\section{Proof of Theorem~\ref{theorem:SCTC}}

\ \ \ \textsc{Theorem} 5.3. (\emph{\textbf{Time and Space Complexities}}).
\emph{Algorithm~\ref{algorithm:skylinecoreness} takes $O(R_{SC} \cdot \Delta_{in} \cdot \Delta_{out})$ time in $O(\Delta \cdot \min\{\Delta_{in}, \Delta_{out}\})$ space. The total time and space complexities for  computing all vertices' corenesses are $O(R_{SC} \cdot \Delta_{in}  \cdot m)$ and $O(\min\{\Delta_{in}, \Delta_{out}\} \cdot  m)$, respectively.}

\begin{proof}
The time complexity of Algorithm~\ref{algorithm:skylinecoreness} is dominated by the iterative computation of $n$-order D-index. In each round, Algorithm~\ref{algorithm:Dindexcomputation} examines at most $deg^{in}_G(v) \cdot deg^{out}_G(v)$ pairs. Therefore, the time complexity of Algorithm~\ref{algorithm:skylinecoreness} is $O(R_{SC} \cdot deg^{in}_G(v) \cdot deg^{out}_G(v))$ $=$ $O(R_{SC} \cdot \Delta_{in} \cdot \Delta_{out})$.  The total time complexity for  all vertices is $O(\sum \limits_{v\in V_G} R_{SC} \cdot deg^{in}_G(v) \cdot deg^{out}_G(v))$ $=$ $O(R_{SC} \cdot \Delta_{in} \cdot \sum \limits_{v\in V_G}  deg_G(v))$ $=$ $O(R_{SC} \cdot \Delta_{in}  \cdot m)$.

Next, we analyze the space complexity. 
To compute $v$'s $n$-order D-index, Algorithm~\ref{algorithm:skylinecoreness} stores the $(n-1)$-order D-indexes of all $v$'s neighbors. The size of vertex $v$'s $n$-order D-index is bounded by $\min\{\Delta_{in}, \Delta_{out}\}$. Thus, the space complexity of  Algorithm~\ref{algorithm:skylinecoreness} is $O(deg_G(v)\cdot \min\{\Delta_{in}, \Delta_{out}\})$ $=$ $O(\Delta \cdot \min\{\Delta_{in}, \Delta_{out}\})$. The total time complexity for  all vertices is $O(\sum \limits_{v\in V_G}  \min\{deg^{in}_G(v), deg^{out}_G(v)\} \cdot deg^{out}_G(v))$ $=$ $O(\min\{\Delta_{in}, \Delta_{out}\} \cdot \sum \limits_{v\in V_G} deg_G(v))$ $=$ $O(\min\{\Delta_{in}, \Delta_{out}\} \cdot  m)$.
\end{proof}

\section{Proof of Theorem~\ref{theorem:SCMC}}

\ \ \ \textsc{Theorem} 5.4. (\emph{\textbf{Message Complexity}}).
\emph{The message complexity of Algorithm~\ref{algorithm:skylinecoreness} is $O(\Delta^2)$. The total message complexity for computing all vertices' corenesses is  $O(\Delta \cdot  m)$. }

\begin{proof}


The message complexity of  Algorithm~\ref{algorithm:skylinecoreness} is dominated by the computation of $n$-order D-index for vertex $v$. When the $n$-order D-index changes from the $(n-1)$-order D-index, $v$ broadcasts messages to its neighbors. The $n$-order D-index contains skyline pairs. Assume that in each round the skyline pair decreases one dimension by one at most. There are a total of $deg_G(v)$ rounds for $v$. Hence, the  message complexity of Algorithm~\ref{algorithm:skylinecoreness}  is $O(deg_G(v) \cdot deg_G(v))$ $=$ $O(\Delta^2)$.  The total message complexity for all vertices is $O(\sum \limits_{v \in V_G} deg_G(v) \cdot deg_G(v))$ $=$ $O(\Delta \cdot \sum \limits_{v \in V_G} deg_G(v))$ $=$ $O(\Delta \cdot m)$.
\end{proof}

\section{Performance comparisons on a single machine}
\begin{figure}[!t]
\centering
\setlength{\abovecaptionskip}{-2pt}
\includegraphics[width = 0.4\textwidth]{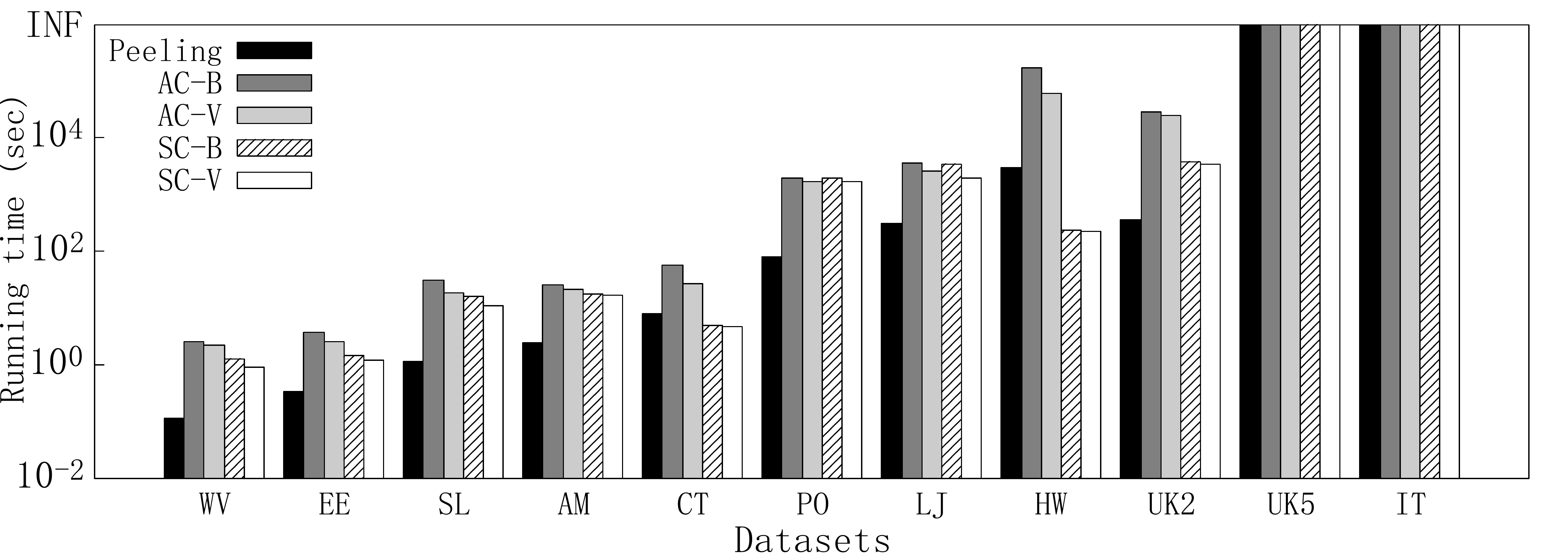}
\caption{Performance comparisons on a single machine}
\label{fig:exp4-revision}
\end{figure}

This set of experiments evaluates the performance of our proposed algorithms and the state-of-the-art centralized method~\cite{dcore-cs} over a single machine. Figure~\ref{fig:exp4-revision} shows the running time results. 
We can observe that for most of the small graphs, the peeling algorithm is more efficient than our algorithms. The reason is two-folded:
(1) our proposed algorithms require iterative computations of in-H-index/out-H-index/D-index for every vertex before convergence, which is time consuming in centralized settings;
(2) on a single machine, the vertex state update is transparent to other vertices in the peeling algorithm while it 
needs message communication exchanges in our algorithms, which incur a higher overhead.
Nevertheless, our proposed algorithms run much faster than the peeling method in distributed settings as reported in Section~\ref{exp:efficiency}. 
On the other hand, for billion-scale graphs \texttt{UK5} and \texttt{IT}, all algorithms could not finish, due to memory overflows. This indeed motivates the need of distributed D-core decomposition over large directed graphs using a collection of machines.  

\end{document}